\documentclass[journal,12pt,onecolumn,draftclsnofoot,romanappendices]{IEEEtran}

\def\paperversion{2} %

\usepackage{todonotes}

\IEEEoverridecommandlockouts
\usepackage{cite}
\usepackage{url}
\usepackage{enumerate}

\usepackage[T1]{fontenc}

\usepackage{booktabs}
\usepackage{multirow}
\usepackage{subcaption}

\usepackage[font=footnotesize]{caption}
\usepackage[font=footnotesize]{subcaption}

\usepackage[cmex10]{amsmath}
\usepackage{amssymb}
\usepackage{amsthm}
\usepackage{graphicx}

\usepackage[ruled,vlined]{algorithm2e}
\SetKw{Continue}{continue}

\usepackage{tikz}
\usetikzlibrary{matrix,arrows,shapes,positioning,chains,scopes,decorations.markings,calc,patterns}
\tikzset{->-/.style={decoration={markings,mark=at position .5 with {\arrow{>}}},postaction={decorate}}}

\tikzset{endblk/.style={
    rounded rectangle,minimum size=6mm,
    thick, draw,%
    align=center,midway,
    font=\small}
    }
\tikzset{process/.style={
    rectangle,minimum size=6mm,
    thick, draw,%
    align=center,midway,
    font=\small}
}
\tikzset{conditional/.style={
    shape aspect=3,rounded corners=2mm,
    diamond,minimum size=6mm,
    thick, draw,%
    align=center,midway,
    font=\small}
}

\tikzset{skip loop/.style={to path={-- ++(0,#1) |- (\tikztotarget)}}}

\tikzset{point/.style={coordinate},>=stealth',draw=black!70,
  arrow/.style={->},every join/.style={rounded corners},
  hv path/.style={to path={-| (\tikztotarget)}},
  vh path/.style={to path={|- (\tikztotarget)}},
  lyes/.style={label=177:yes},
  lno/.style={label=177:no},
  ryes/.style={label=3:yes},
  rno/.style={label=3:no},
  bno/.style={label=-93:no},
  byes/.style={label=-93:yes},
}

\usepackage{pgfplots}
\usepackage{pgfplotstable}
\usepackage{booktabs}
\usepackage{colortbl}

\makeatletter
\newcommand{\removelatexerror}{\let\@latex@error\@gobble}
\makeatother

\newtheorem{lemma}{Lemma}
\newtheorem{theorem}{Theorem}
\newtheorem{corollary}{Corollary}
\theoremstyle{definition}\newtheorem{definition}{Definition}

\DeclareMathOperator{\diag}{diag}

\DeclareMathOperator{\rk}{rk}

\DeclareMathOperator*{\argmin}{arg\,min}
\DeclareMathOperator*{\argmax}{arg\,max}
\DeclareMathOperator{\Span}{span}
\DeclareMathOperator{\dom}{dom}

\newcommand*{\Scale}[2][4]{\scalebox{#1}{$#2$}}%

\begin{document}

\title{A Unified Adaptive Recoding Framework for Batched Network Coding}

\author{\IEEEauthorblockN{
	Hoover~H.~F.~Yin, Bin~Tang, Ka~Hei~Ng, Shenghao~Yang, Xishi~Wang, and Qiaoqiao~Zhou
}
	\thanks{This paper was presented in part at 2019 IEEE International Symposium on Information Theory \cite{uni}. %
	}
	\thanks{H.~Yin is with the n-hop technologies Limited, Hong Kong, China and the Institute of Network Coding, The Chinese University of Hong Kong, Hong Kong, China.
	B.~Tang is with the School of Computer and Information, Hohai University, Nanjing, China.
	K.~Ng is with the Department of Physics, The Chinese University of Hong Kong, Hong Kong, China.
	S.~Yang is with the School of Science and Engineering, The Chinese University of Hong Kong, Shenzhen, Shenzhen, China.
	He is also with Shenzhen Key Laboratory of IoT Intelligent Systems and Wireless Network Technology and Shenzhen Research Institute of Big Data, Shenzhen, China.
	X.~Wang and Q.~Zhou are with the Department of Information Engineering, The Chinese University of Hong Kong, Hong Kong, China.
	Emails: \mbox{hfyin@inc.cuhk.edu.hk}, \mbox{cstb@hhu.edu.cn}, \mbox{kaheicanaan@link.cuhk.edu.hk}, \mbox{shyang@cuhk.edu.cn}, \mbox{wx116@ie.cuhk.edu.hk}, \mbox{zq115@ie.cuhk.edu.hk}
	}
	\thanks{This work was funded in part by the Shenzhen Science and Technology Innovation Committee (Grant JCYJ20180508162604311, ZDSYS20170725140921348).}
}

\maketitle

\begin{abstract}
Batched network coding is a variation of random linear network coding which has low computational and storage costs.
In order to adapt to random fluctuations in the number of erasures in individual batches, it is not optimal to recode and transmit the same number of packets for all batches.
Different distributed optimization models, which are called adaptive recoding schemes, were formulated for this purpose.
The key component of these optimization problems is the expected value of the rank distribution of a batch at the next network node, which is also known as the expected rank.
In this paper, we put forth a unified adaptive recoding framework with an arbitrary recoding field size.
We show that the expected rank functions are concave when the packet loss pattern is a stationary stochastic process, which covers but not limited to independent packet loss and Gilbert-Elliott packet loss model.
Under this concavity assumption, we show that there always exists a solution which not only can minimize the randomness on the number of recoded packets but also can tolerate rank distribution errors due to inaccurate measurements or limited precision of the machine.
We provide an algorithm to obtain such an optimal optimal solution, and propose tuning schemes that can turn any feasible solution into a desired optimal solution.
\end{abstract}

\section{Introduction}

It is well-known that for a wide range of scenarios, random linear network coding (RLNC) \cite{random,random2,jaggi03,Sanders03}, which is a simple realization of network coding \cite{flow,alg,linear}, can achieve the capacity of networks with packet loss \cite{rate,Lun2008}.
The major challenges for practical implementation of RLNC include the high computational and storage costs of network coding at the intermediate nodes, and the high coefficient vector overhead.
\emph{Batched network coding} \cite{chou03,Silva2009,Heidarzadeh2010,Mahdaviani12,yang14bats} provides a general framework for resolving these issues by allowing network coding only on a relatively small number of packets. 
A batched network code consists of an inner code and an outer code.
The outer code encodes the packets for transmission into coded packets that are partitioned into small subsets, called \emph{batches}.
The inner code is formed by \emph{recoding}, which applies RLNC to the packets belonging to the same batch.

Batched network coding includes many RLNC schemes as special cases.
Early works used disjoint batches \cite{chou03}, which is also called generation based RLNC.
When there is only a single batch, BNC becomes the generation-based RLNC or its variations~\cite{lucani18fulcrum,nguyen20dsep} which use binary field in the inner code to reduce the computational cost.
Batches with certain overlapping structure has been studied in \cite{Silva2009,Heidarzadeh2010,yaoli11,bin_expander15}, where the outer code involves only repetition and partition.
More advanced outer codes can be designed by extending fountain codes \cite{yang14bats} and LDPC codes \cite{Mahdaviani12,Mahdaviani13,bin18ldpc}. 

For a general batched network coding, the end-to-end network operations on a batch can be modelled by a \emph{batch transfer matrix}.
The rank of the transfer matrix of a batch is also called the rank of the batch.
The achievable rate of a batched network code is upper bounded by the expected rank of the batches \cite{yang11x2}.
There exist batched network codes which have close-to-optimal achievable rate and low encoding and decoding complexities, e.g., BATS codes \cite{yang14bats}.

The number of recoded packets to be generated for a batch is called the \emph{recoding number} of this batch.
The recoding number problem, i.e., determining the recoding numbers of the batches at a network node, is a core of inner code design which can affect the expected rank of the batches and thus the achievable rate.
Due to the ease of analysis and simplicity, \emph{baseline recoding}, which assigns the same recoding number for all batches, was adopted in many literature such as \cite{variable,fun2,pro2,zhou17b,bats_schedule,delay,buffer}.
When the number of packets generated by the outer code for a batch (the \emph{batch size}) tends to infinity, baseline recoding asymptotically achieves the optimal throughput.
However, it is not the case for finite batch sizes \cite{yang14a}.
A recoding scheme that allows different recoding numbers for different batches is called \emph{adaptive recoding} \nocite{rs}\cite{scheduling,adaptive}.
It was shown in \cite{scheduling} that assigning recoding numbers randomly, i.e., random scheduling \cite{rs}, is not an optimal adaptive recoding approach.
By designing specific optimization models, adaptive recoding can outperform baseline recoding and also provide flexibility for practical implementations. %

In literature, the recoding number problem has been investigated using different formulations with different packet loss models and field size assumptions \cite{adaptive,scheduling,wang2021smallsample,yin21impact,intrablock,ge_adaptive,rf,zhang2017efficient}.
\begin{itemize}
	\item The independent packet loss model is employed in \cite{adaptive,scheduling,wang2021smallsample,yin21impact}, and the Gilbert-Elliott model \cite{GilbertBurst,ElliottBurst} for burst loss is employed in \cite{yin21impact,ge_adaptive,intrablock}.
	\item An arbitrary finite field size is allowed in \cite{scheduling,rf}, while the field size in \cite{adaptive,ge_adaptive,zhang2017efficient} tends to infinity.
	\item An integer programming problem for optimizing the recoding number of each batch is employed in \cite{adaptive,rf,yin21impact}.
		A probabilistic approach to avoid integer programming is employed in \cite{scheduling,ge_adaptive,zhang2017efficient}, which optimizes the probability distribution of the recoding number conditioning on the rank of the batch (with an artificial upper bound on the recoding numbers) by linear programming.
	\item The models in \cite{scheduling,ge_adaptive,rf,zhang2017efficient} consider a known rank distribution of all the batches, while those in \cite{adaptive,yin21impact,wang2021smallsample,intrablock} sample the rank distribution in run-time.
\end{itemize}
Adaptive recoding is a general recoding framework that can be used in various related research including network utility maximization \cite{dong20}, rank distribution inference \cite{wang2021smallsample}, interleaving \cite{intrablock} and relaying with overhearing \cite{rf,zhang2017efficient}.
Therefore, a unified theoretical framework for recoding number optimization is desired.

In this paper, we propose a unified adaptive recoding framework to study the recoding number problem with an arbitrary finite field size.
The packet loss pattern on a link can be a general stochastic process subject to a technical condition that the expected rank functions used in the problem formulation are concave.
We show that a stationary packet loss pattern, including both the independent loss pattern and the burst packet loss pattern modelled by a time-homogeneous Markov chain as special cases, satisfies this concavity requirement.
Given the rank distribution of the batches at a node, we study a general recoding number optimization problem that maximizes the average expected rank at the next network node. %

We show that our problem has a special optimizer where for at most one rank values, the recoding number is a ``non-integer'', which is also called an \emph{almost deterministic solution (ADS)}. 
A non-integer recoding number can be used in a probabilistic manner: The recoding number is selected from two consecutive integers randomly, or a group of batches of the same rank is separated into two subgroups where each of which uses one of the consecutive integer recoding numbers. 
An ADS not only can solve the linear programming adaptive recoding models, but also can be modified to a solution for the integer programming adaptive recoding models.

As our problem is a concave optimization problem, a general optimization solver can solve it, but such a solver may not give an ADS.
We propose a coordinate decent approach to solve our problem efficiently, benefiting from the piecewise linear structure of the expected rank functions, which always generate an optimal ADS. 
Moreover, the solution generated by our algorithm can tolerate the inaccuracy of the rank distribution: A small error on the rank distribution will not affect too much on the number of recoded packets for the batches, even when the error gives zero probability masses in the rank distribution.
We also provide a way to tune a general primal point, regardless of its optimality, to an optimal ADS with rank distribution inaccuracy tolerance. 
In other words, we can update an existing solution after some parameters of the model or the channel statistics have been changed.
Last but not least, we also present a method to extract such error tolerating optimal ADS from an optimal Lagrange multiplier.

The remainder of this paper is organized as follows.
In Section~\ref{sec:adp}, we give a brief description on batched network coding and formulate the unified adaptive recoding framework.
The main results of this paper are also summarized. 
The main technical parts are presented in Sections~\ref{sec:exp}, \ref{sec:uni} and \ref{sec:tune}, where we discuss the expected rank functions, properties of the framework and the algorithms respectively. Concluding remarks are in Section~\ref{sec:conclude}.

\section{General Adaptive Recoding} \label{sec:adp}

In this section, we first give a brief introduction to batched network coding.
Then, we model the unified adaptive recoding framework from scratch which generalizes the recoding number optimization problems in \cite{scheduling,adaptive}.

\subsection{Batched Network Coding}

Suppose we are transmitting a file from a source node to a destination node in a network.
The file consists of a number of \emph{input packets}, each of which is regarded as a vector in $\mathbb{F}_q$, the finite field of $q$ elements. 
To apply a batched network code, the source node runs an outer code encoder to generate \emph{batches}, each of which consists of $M$ \emph{coded packets}, where $M$ is also known as the \emph{batch size}.
A coded packet is a linear combination of the input packets.
We can choose, for example, an outer code introduced in \cite{yang14bats,bin_expander15,bin18ldpc}. 

Without loss of generality, here we assume that recoding uses the same field $\mathbb{F}_q$.
In general, a subfield of $\mathbb{F}_q$ can be used for recoding as discussed in \cite{protocol,bats_book,lucani18fulcrum}.
A coefficient vector of $M$ symbols from $\mathbb{F}_q$ is attached to each coded packet, and the coefficient vectors of all the $M$ packets of a batch freshly generated by an encoder form an identity matrix.

Consider \emph{recoding} at a network node that is not the destination node.
If the node is the source node, it receives batches from the outer code encoder.
If the node is an intermediate network node, it receives the batches from other network nodes on its upper stream.
The node will transmit only recoded packets generated by random linear combinations of the packets belong to the same batch, where the coefficients of the random linear combinations are chosen uniformly at random from $\mathbb{F}_q$. 
The recoded packets are supposed to be of the same batch.
We will discuss how to determine the number of recoded packets to transmit for each batch in the next subsection.

\begin{definition}[Rank]
Two packets of a batch are called \emph{linearly independent} if their coefficient vectors are linearly independent.
For all the received packets of a batch, the rank of the matrix formed by the coefficient vectors of these packets is also called the \emph{rank} of the batch.
\end{definition}

It has been shown that the ranks of the batches received at a node form a sufficient statistic to the performance of the inner code~\cite{bats_book}.
Consider the transmission of a batch on a single path from the source node to the destination node.
The rank of the batch at each node on the path is a random variable depending on the packet loss pattern.
A batch is of rank $M$ at the source node.
As packet loss cannot increase the rank, with probability $1$, the rank of a batch at a node on the path is no less than the rank of the batch at the next node on the path.
The rank distribution of the batch at the destination node determines the performance of the batched network code.

The destination node collects batches for decoding the input packets.
A necessary condition for successful decoding is that the total rank of all the batches is at least the number of input packets.
For the outer codes introduced in \cite{yang14bats,bin_expander15,bin18ldpc}, a belief propagation (BP) algorithm can be applied to decode the batches efficiently, and achieves a rate very close to the average rank of all the batches used for decoding.
BP decoding, however, requires a large number of batches to have a high successful probability.
When the number of batches is small, e.g., 10 to 100, BP decoding may stop with a high probability before decoding a required fraction of input packets.
When BP decoding stops, the inactivation decoding can continue the decoding and achieves nearly optimal decoding performance \cite{Raptormono,inactivation}.

\subsection{Recoding Number Optimization}

Here we discuss a general distributed model for optimizing the recoding numbers from scratch.
For the sake of readability, we recall the definition of recoding number.

\begin{definition}[Recoding Number]
	The \emph{recoding number} of a batch at a network node is the number of recoded packets generated for this batch.
\end{definition}

The recoding numbers should be chosen to maximize the end-to-end expected rank of a batch subject to certain resource constraints to be make clear soon.
This centralized optimization involves all the recoding numbers and link packet loss statistics along the path of a communication, and is usually difficult to be performed in practice.
Therefore, we focus on the distributed model for recoding number optimization that can be done at each network node separately, involving only the statistics from its adjacent nodes and links.
Existing works such as \cite{scheduling} have shown that the distributed optimization approach achieves a close-to-optimal performance.

As a node-by-node distributed model, we only need to consider the link where the packets come from and the link where the recoded packets send to.
That is, we do not concern about the overall network topology.
On the other hand, we suppose each node has the required statistics from its adjacent nodes and links before optimizing the recoding numbers.
These statistics can be inaccurate and may be updated from time to time.
We will discuss the way to adapt to inaccurate statistics in Section~\ref{sec:preferred} and updated statistics (or other parameters) in Section~\ref{sec:param}.
In other words, our model can support different designs for obtaining the statistics, controlling the network flow and buffer size, determining the parameters of the batched network code, etc.
This also means that our model can work with or without feedback, as the effect of feedback in the view of our model is the changes of the statistics and parameters.
As a quick summary, we assume that the parameters involved in our model are fed by other network protocols, and we only concern how to optimize the recoding numbers under the given parameters.

Now, we consider recoding at a network node $V$ where the recoded packets at this node are transmitted to the next node $V^+$.
Here $V$ can be the source node or an intermediate node but not the destination node, which is because recoding at the destination node gives no help to the performance of decoding.
There is a packet erasure channel from $V$ to $V^+$ where the packet loss pattern forms a stochastic process.
It is sufficient for us to consider stationary loss pattern for the time-being (see further discussion in Section~\ref{sec:exp:1}).
Note that both the independent loss pattern and the burst packet loss pattern modelled by a time-homogeneous Markov chain are stationary.

\begin{definition}[Input Rank Distribution]
	The \emph{input rank distribution} at a network node is the rank distribution of the received batches at this node.
\end{definition}

Denote by $(h_0, h_1, \ldots, h_M)$ the input rank distribution at $V$.
If $V$ is the source node, then we have $h_M = 1$.
When a batch of rank $r$ is received, the node will transmit $t$ recoded packets of this batch with probability $\alpha_{t|r}$, which is the main object to be studied in this paper. 
The average recoding number of the batches, denoted by $t_\text{avg}$, specifies the network resource allocated to transmit these batches, which is related to the flow control, the buffer stability, etc.
The value of $t_\text{avg}$ can be determined by resource allocating~\cite{dong20}, which is out of the scope of this paper.
We assume that $t_\text{avg}$ is given, and we call this value the \emph{total resource}.

Our objective to design $\alpha_{t|r}$ is to maximize the (average) expected rank of the batches at the next network node $V^+$.
For a batch of rank $r$ at V, let $E_r(t)$ be the expected rank of this batch at $V^+$ when $t$ recoded packets of this batch are transmitted.
The formula of $E_r(t)$ can be obtained based on the packet loss pattern (see Section~\ref{sec:exp:1}). 
We must have $E_r(0) = 0 \le E_r(t) \le r$ for all non-negative integers $r$ and $t$, because
\begin{enumerate}[i)]
	\item the next node must receive no packet if no packet is transmitted at the current node, i.e., the expected rank at $V^+$ is $0$; and
	\item the linear combinations of $r$ independent vectors cannot result in more than $r$ independent vectors.
\end{enumerate}
Moreover, we must have $E_0(t) = 0$ for all non-negative integers $t$.
In practice, we have a batch of rank $0$ when the whole batch is lost or only zero packets of this batch are received.
We consider that for each $r$, $E_r(t)$ is a monotonic increasing concave function.
This is because a newly received packet of a batch
\begin{itemize}
	\item is either linearly dependent or independent of the already received packets of the same batch, i.e., the rank of a batch cannot be dropped by receiving a new packet; and
	\item has a non-increasing chance of being linearly independent of the already received packets of the same batch when there are more and more received packets of this batch.
\end{itemize}
We will show that when the packet loss pattern is stationary, $E_r(t)$ is a monotonic increasing concave function (see Theorem~\ref{thm:concave}).
Note that there may have non-stationary packet loss pattern which gives concave expected rank functions (see Section~\ref{sec:eg_non_stat}).

Denote by $\mathbb{N}$ the set of non-negative integers.
Define $[n] = \{0, 1, \ldots, n\}$ for any $n \in \mathbb{N}$.
We formulate the following recoding number optimization problem:
\begin{equation}
\tag{P} \label{eq:P}
\begin{IEEEeqnarraybox}[][c]{rCl}
	\max_{0 \le \alpha_{t|r} \le 1, \forall r \in [M], t \in \mathbb{N}} & \quad & \sum_{r = 0}^M h_r \sum_{t = 0}^\infty \alpha_{t|r} E_r(t)\\
	\mathrm{s.t.} && \sum_{r = 0}^M h_r \sum_{t = 0}^\infty t \alpha_{t|r} = t_\text{avg}, \\
				  && \sum_{t = 0}^\infty \alpha_{t|r} = 1, \forall r \in [M].
\end{IEEEeqnarraybox}
\end{equation}
The objective of \eqref{eq:P} is the average expected rank at $V^+$. 
The first constraint in \eqref{eq:P} is to ensure that we fully utilize the allocated total resource, which implicitly assumes that the expectation of the distribution $\{\alpha_{t|r}\}_{t = 0}^\infty$ for each $r \in [M]$ converges.
The second constraint in \eqref{eq:P} states that for every $r \in [M]$, $\{\alpha_{t|r}\}_{t = 0}^\infty$ is a probability distribution.
Although $\alpha_{t|0}$ for all $t > 0$ are variables in \eqref{eq:P}, it is safe to set $\alpha_{0|0} = 1$ and $\alpha_{t|0} = 0$ for all $t > 0$ as we cannot generate any meaningful recoded packet for a batch of rank $0$.

\subsection{A Unified Adaptive Recoding Framework}

Although the form of \eqref{eq:P} is linear, it has infinitely many variables.
One of our main contributions is that we transform \eqref{eq:P} into the following concave optimization problem with finite variables:
\begin{equation}
  \tag{IP} \label{eq:IP}
  \begin{IEEEeqnarraybox}[][c]{rCl}
    \max_{t_r \ge 0,\forall r \in [M]} & \quad & \sum_{r = 0}^M h_r E_r(t_r)\\
    \mathrm{s.t.} && \sum_{r = 0}^M h_r t_r = t_\text{avg},
  \end{IEEEeqnarraybox}
\end{equation}
where the domain of $E_r(t)$ is extended to non-negative real by linear interpolation, i.e., 
\begin{equation} \label{eq:realE}
	E_r(t) = (t-\lfloor t \rfloor) E_r(\lfloor t \rfloor + 1) + (1 - (t-\lfloor t \rfloor)) E_r(\lfloor t \rfloor).
\end{equation}
We can interpret \eqref{eq:realE} as the expected rank at the next node when the current node transmits $\lfloor t \rfloor$ and $\lfloor t \rfloor + 1$ recoded packets with probabilities $1 - (t - \lfloor t \rfloor)$ and $t - \lfloor t \rfloor$ respectively.

The objective of \eqref{eq:IP} is the one in \eqref{eq:P} after imposing a restriction to the distributions $\{\alpha_{t|r}\}$ for all $r \in [M]$.
That is, this objective is also the average expected rank at $V^+$.
Regarding the restriction, we can see from \eqref{eq:realE} that the support size of each $\{\alpha_{t|r}\}$ is now at most $2$.
Also, if the support size is $2$, then the support consists of two consecutive integers.

In a similar manner, the constraint in \eqref{eq:IP} is the one in \eqref{eq:P} after imposing the same aforementioned restriction to the distributions $\{\alpha_{t|r}\}$ for all $r \in [M]$.
This constraint is to limit the recoding number of each batch.
As it is always no harm or even beneficial to the expected rank at $V^+$ when we send more packets, we know that if we set the constraint as $\sum_{r = 0}^M h_r t_r \le t_\text{avg}$, we can always find an optimal solution such that the equality holds.

We will show in Theorem~\ref{thm:IP} that an optimal solution $\{t_r\}$ of \eqref{eq:IP} induces an optimal solution of \eqref{eq:P}. 
Before going into detail, we first discuss why \eqref{eq:IP} provides a unified framework for adaptive recoding.

One mainstream of adaptive recoding formulation follows the linear programming approach proposed in \cite{scheduling}.
Variations such as \cite{ge_adaptive} substitute different formulations of $E_r(t)$.
These formulations can be regarded as \eqref{eq:P} with an artificial upper bound on the recoding number for all ranks so that the number of variables become finite.
A drawback is that we may not obtain an optimal recoding scheme if we choose an upper bound that is too small.
On the other hand, the number of variables is still huge as each probability mass corresponds to a variable, especially when we choose an upper bound that is too large.
In \eqref{eq:IP}, the issues on the artificial upper bound and the amount of variables are resolved.
The implication of \eqref{eq:IP} is that there is an optimal solution of \eqref{eq:P} such that the size of the support of $\{\alpha_{t|r}\}_{t = 0}^\infty$ is at most $2$ for every rank $r$.
Also, we can relate \eqref{eq:IP} or \eqref{eq:P} to the one with an artificial upper bound $\tilde{M}$ by modifying $E_r(t)$ in a way that $E_r(t) = E_r(\tilde{M})$ for all $t > \tilde{M}$.
Note that the monotonic increasing concave nature of $E_r(t)$ is preserved after the manipulation.

Another mainstream of adaptive recoding formulation follows the integer programming approach proposed in \cite{adaptive}.
This formulation groups a finite number of batches and optimizes the recoding number of each batch in the group, which is later known as blockwise adaptive recoding in \cite{intrablock,yin21impact}.
By symmetry, we should have the same recoding number for the batches of the same rank, except that some batches may send $1$ more recoded packet due to the total resource constraint.
This can be proved by applying Jensen's inequality (see the proof of Theorem~\ref{thm:IP}).
This is, the input rank distribution of \eqref{eq:IP} is the portion of different ranks of the batches in the group.
We will further show in Theorem~\ref{thm:integer} that there exists a solution $\{t_r\}$ solving \eqref{eq:IP} such that there is at most one non-integer $t_r$.
In other words, the recoding number of the batches of the same rank can be differed by $1$ for at most one of the ranks.
This matches the observation of the solution in \cite{adaptive}.

\begin{definition}[ADS]
	An \emph{almost deterministic solution (ADS)} of \eqref{eq:IP} is an optimal solution that except for at most one rank, all recoding numbers are integers.
\end{definition}

\subsection{Preferred Solutions}

The input rank distribution in practice should be known from the statistics and hence may not be accurate or subject to change over time.
Roughly, if the rank distribution changes slightly, the optimal solution may also change.
For robustness, we desire an optimal or a close-to-optimal solution when the input rank distribution is inaccurate.
One concern, for example, is that what if the inaccuracy makes some non-zero masses to zero in the input rank distribution.
For a rank $r$ with $h_r = 0$, any value of $t_r$ does not affect the optimality of \eqref{eq:IP}.
However, when we receive a batch of such rank $r$ in practice, we need to respond reasonably by assigning certain recoding number to this batch.

On the other hand, an ADS can further reduce the randomness of the recoding number which can simplify the analysis and design of other applications based on adaptive recoding.
Therefore, if possible, we prefer an ADS which at the same time is robust against the inaccuracy of input rank distribution.
We will formally define 
preferred solution in Section~\ref{sec:preferred}.  

\begin{definition}[(Informal) Preferred Solution] \label{def:preferred}
	A \emph{preferred solution} of \eqref{eq:IP} is an optimal ADS which can tolerate rank distribution errors.
\end{definition}

As a concave optimization problem, any general convex optimization solver can be applied to solve \eqref{eq:IP}.
However, the solution may not be a preferred solution.
So, we derive an algorithm for solving \eqref{eq:IP} by analyzing the properties of the problem (see details in Sections~\ref{sec:2f}, \ref{sec:preferred} and \ref{sec:primal}).
From the piecewise linear structure of $E_r(t)$, we know that the slope of the line joining $(t, E_r(t))$ and $(t+1, E_r(t+1))$ is
\begin{equation}\label{eq:Delta}
  \Delta_{r,t} := E_r(t+1) - E_r(t).
\end{equation}
Due to the special properties of \eqref{eq:IP}, we derive a coordinate decent algorithm in Algorithm~\ref{alg:opt}, where an interior point $\{t_r\}_{r = 0}^M$ means that $\sum_r t_r < t_\text{avg}$ and $t_r \ge 0$ for all $r \in [M]$.
The algorithm does not guarantee a preferred solution unless we start from an interior point with certain properties (see Theorem~\ref{thm:opt}).
One working example is that we run Algorithm~\ref{alg:opt} with $t_r = 0$ for all $r \in [M]$ and $u = t_\text{avg}$. %

\begin{figure}
\removelatexerror
\begin{algorithm}[H]
  \footnotesize
  \caption{Searching from an interior point}
  \label{alg:opt}
  \KwData{An interior point $\{t_r\}_{r = 0}^M$, the remaining resource $u$, the input rank distribution $(h_0,\ldots,h_M)$, and the access to the values $\Delta_{r,t}$.}
  \KwResult{A feasible solution $\{t_r\}_{r = 0}^M$.}
  \While{$u > 0$}{
    $r \leftarrow$ an element in $\argmax_{r \in [M]} \Delta_{r,\lfloor t_r \rfloor}$ \;
    \If{$h_r (1-(t_r - \lfloor t_r \rfloor)) \le u$}{
      $u \leftarrow u - h_r (1-(t_r - \lfloor t_r \rfloor))$ ;
      $t_r \leftarrow \lfloor t_r \rfloor + 1$ \;
    }\Else{
      $t_r \leftarrow t_r + u/h_r$ ;
      $u \leftarrow 0$ \;
    }
  }
  \Return $\{t_r\}_{r = 0}^M$ \;
\end{algorithm}
\end{figure}

In a nutshell, Algorithm~\ref{alg:opt} is a greedy algorithm which climbs the steepest slope in each iteration.
Due to the piecewise linear structure, we can increase the recoding number for a rank by $1$ if there is enough resource to do so.
We also have an algorithm to tune any feasible solution, regardless of its optimality, to a preferred solution, but we defer its discussion to Section~\ref{sec:primal2}.
The time complexities of these algorithms will be discussed in Sections~\ref{sec:primal} and \ref{sec:primal2}.
By making use of these algorithms, we can further tune an arbitrary primal point to a preferred solution (see Section~\ref{sec:param}), which can be applied to tune an existing solution to the one after updating the parameters of the model, e.g., updating input rank distribution.
Besides, we can also tune a preferred solution from an optimal Lagrange multiplier (see Section~\ref{sec:dual}), which is useful when a dual-based solver is used.

We now use an example to demonstrate the error tolerating property of a preferred solution.
Suppose $M = 4$, $q \to \infty$, $t_\text{avg} = 4$ and an input rank distribution $\mathbf{h}_1 = (0.0625,$ $0.25,$ $0.375,$ $0.25,$ $0.0625)$.
Also, suppose the outgoing channel an independent packet loss channel with $20\%$ loss rate.
The optimal recoding numbers $(t_0, t_1, \ldots, t_4)$ are $(0, 2.25, 4, 6, 7)$.
Now, let $\mathbf{h}_2 = (0.0625,$ $0.2,$ $0.425,$ $0.3125,$ $0)$ be the input rank distribution we measured.
An arbitrary solver may obtain the recoding numbers in the form of $(0, 0.2145, 4, 6, X)$ for an arbitrary $X$ as the value of $X$ does not affect the objective value.
However, when we receive a batch of rank $4$, we need to choose a reasonable $X$.
If we solve \eqref{eq:IP} by Algorithm~\ref{alg:opt}, we obtain the optimal recoding numbers $(0, 0.2145, 4, 6, 7)$.
We can observe that most of the recoding numbers are the same when we switch between $\mathbf{h}_1$ and $\mathbf{h}_2$, i.e., the solution can tolerate small changes in the input rank distribution.
Also, our algorithm ensures that we choose a reasonable recoding number for those zero masses in the input rank distribution.

\section{Expected Rank Functions} \label{sec:exp}

We can see in our recoding number optimization problem \eqref{eq:P} that the expected rank functions $E_r(t)$, which model the channels, are the key component.
In this section, we formulate the expected rank functions mathematically and present their properties.

\subsection{Formulation}\label{sec:exp:1}

Let $\{Z_t\}$ be the packet loss pattern induced by the outgoing channel, which is a binary stochastic process (a sequence of binary random variables) where
$Z_t = 1$ if the $t$-th packet is received at the next network node, or $Z_t = 0$ otherwise.
Without loss of generality, let $Z_1$ be the random variable for the first transmitted packet of a batch.
As the recoded packets are generated by RLNC over $\mathbb{F}_q$, the expected rank function can be expressed by
\begin{equation} \label{eq:ert}
	E_r(t) = \sum_{i = 0}^t \Pr \left( \sum_{j = 1}^t Z_j = i \right) \sum_{j = 0}^{\min\{i,r\}} j \zeta_j^{i,r},
\end{equation}
where $\zeta_j^{i,r}$ is the probability that a batch of rank $r$ at the current node with $i$ received packets at the next node has rank $j$ at the next node.
The exact formulation of $\zeta_j^{i,r}$ can be found in \cite{yang14bats}, which is
$\zeta_j^{i,r} = \frac{\zeta_j^i \zeta_j^r}{\zeta_j^j q^{(i-j)(r-j)}}$, where $\zeta_j^m = \prod_{k = 0}^{j-1} (1-q^{-m+k})$.

The complicated formula of $\zeta_j^{i,r}$ makes the analysis difficult.
Instead of analysing \eqref{eq:ert} directly, we consider another formulation by using random matrices.
The rank of a matrix $\mathbf{M}$ is denoted by $\rk(\mathbf{M})$.
A $y \times x$ matrix $\mathbf{M}$ where $y = 0$ or $x = 0$ is an empty matrix and $\rk(\mathbf{M})=0$.
Denote by $\mathbb{E}[\cdot]$ the expectation operator.
Let $\mathbf{R}_{r,t}$ be an $r \times t$ totally random matrix with entries distributed uniformly over $\mathbb{F}_q$.
Define a $t \times t$ diagonal matrix $\mathbf{Z}_t$ over $\mathbb{F}_q$, where
	$\mathbf{Z}_t := \diag(Z_1, Z_2, \ldots, Z_t)$.

\begin{lemma} \label{lem:ert_matrix}
	$E_r(t) = \mathbb{E}[\rk(\mathbf{R}_{r,t}\mathbf{Z}_t)]$.
\end{lemma}

\begin{IEEEproof}
	See Appendix~\ref{sec:lem:ert_matrix}.
\end{IEEEproof}

\subsection{Sufficient Condition for the Concavity}

Let $\dom(g)$ denote the domain of a function $g$. For an integer $t$ and any real-valued function $g$ where $\{t-1, t, t+1\} \subseteq \dom(g) \subseteq \mathbb{Z}$, we say $g$ is \emph{concave} at $t$ if and only if
\begin{equation} \label{eq:concaveg}
	g(t+1) - g(t) \le g(t) - g(t-1).
\end{equation}
The concavity is \emph{strict} at $t$ if and only if the inequality in \eqref{eq:concaveg} is strict.
When the left side of \eqref{eq:concaveg} equals its right side, we call $g$ is \emph{linear} at $t$.
When we do not specify the point $t$, then we say $g$ is concave if and only if $g$ is concave at all $t$ such that $\{t-1, t, t+1\} \subseteq \dom(g)$.
We also define strictly concave and linear in a similar manner.
On the other hand, we say $g$ is \emph{monotonic (or strictly) increasing} at $t$ if and only if $g(t+1) \geq \!\!(\text{or}>)\; g(t)$.
We do not specify the point $t$ when the increment is true for all $\{t, t+1\} \subseteq \dom(g)$.

Before we discuss the concavity of $E_r(t)$, let us first show some properties of $\mathbb{E}[\rk(\mathbf{R}_{r,t})]$ in the following lemma.
To simplify the notations, define 
	$\Delta \mathbb{E}[\rk(\mathbf{R}_{r,t})] := \mathbb{E}[\rk(\mathbf{R}_{r,t+1})] - \mathbb{E}[\rk(\mathbf{R}_{r,t})]$.

\begin{lemma} \label{lem:concaveR}
	$\mathbb{E}[\rk(\mathbf{R}_{r,t})]$ has the following properties:
	\begin{enumerate}[a)]
		\item $\mathbb{E}[\rk(\mathbf{R}_{r,t})]$ is monotonic increasing and concave with respect to $t$. %
			The concavity is strict if $r > 0$ and $q$ is finite.
			When $q \to \infty$, the concavity at $r$ is strict if $r > 0$; \label{lem:concaveR:a}
		\item $\Delta \mathbb{E}[\rk(\mathbf{R}_{r,t})]$ is monotonic increasing with respect to $r$.
			The increment is strict when $q$ is finite.
			When $q \to \infty$, the increment is strict at $t$. \label{lem:concaveR:b}
	\end{enumerate}
\end{lemma}

\begin{IEEEproof}
	See Appendix~\ref{sec:lem:concaveR}.
\end{IEEEproof}

Most commonly used channel models are stationary stochastic processes.
For example, the independent packet loss model is a Bernoulli process, which is stationary.
For burst packet loss channels, the Gilbert-Elliott (GE) model \cite{GilbertBurst,ElliottBurst}, which is a $2$-state hidden Markov model, is widely adopted in various applications \cite{gb_eg1,gb_eg2,gb_eg3,frohn11}.
The transition probabilities of the GE model can be estimated by the average burst error length and the average number of packet drops \cite{hasslinger08}, or be trained by the Baum-Welch algorithm \cite{hmm}.
A higher number of states can model the channel more accurately \cite{morestate,nstate,4state}, which can be used to capture physical properties like BPSK coding with Rayleigh fading process \cite{wang95,ge_pkt}.
These Markov chain models are time-homogeneous, %
i.e., the models are stationary.
Although there may be multiple internal states in a channel model, its packet loss pattern is stationary if the model is stationary \cite{protocol}.
The following theorem states that a stationary packet loss pattern is a sufficient condition for the concavity of $E_r(t)$, which is one of the main results in this paper.
In other words, it is sufficient to consider concave $E_r(t)$ in most scenarios.
Define
	$\mathbf{Y}_t := \diag(Z_2, \ldots, Z_t)$.

\begin{theorem} \label{thm:concave}
	If $\{Z_t\}$ is a stationary stochastic process, then
	$E_r(t)$ is a monotonic increasing concave function.
	Furthermore, $E_r(t)$ is strictly concave at $c \ge 1$ for all $r > 0$ if
	\begin{enumerate}[i)]
		\item $q$ is finite and $\Pr(Z_1 = 1, Z_{c+1} = 1) \neq 0$; or
		\item $q \to \infty$ and $\Pr(Z_1 = 1, Z_{c+1} = 1, \rk(\mathbf{Y}_c) = r-1) \neq 0$.
	\end{enumerate}
\end{theorem}

\begin{IEEEproof}
	See Appendix~\ref{sec:thm:concave}.
\end{IEEEproof}

\begin{corollary} \label{cor:linear}
	If $\{Z_t\}$ is a stationary stochastic process and $q \to \infty$, then $E_r(t)$ is linear at $t < r$.
\end{corollary}

\begin{IEEEproof}
	By Theorem~\ref{thm:concave}, we know that $E_r(t)$ is concave.
	Also, the concavity is strict at $c$ if $\Pr(Z_1 = 1, Z_{c+1} = 1, \rk(\mathbf{Y}_c) = r-1) \neq 0$.
	However, this condition is not possible when $c < r$, which means that $E_r(t)$ is linear at $t < r$.
\end{IEEEproof}

\subsection{Example of Concavity with a Non-Stationary Process}
\label{sec:eg_non_stat}

Note that there exist non-stationary stochastic processes which can result a concave $E_r(t)$.
We give a simple example here.
Let $Z_1 = 1$ and $\{Z_2, Z_3, \ldots\}$ is a Bernoulli process with packet loss rate $p < 1$.
Further, let $q \to \infty$.
Although the stochastic process $\{Z_1, Z_2, \ldots\}$ is not stationary, the sub-process $\{Z_2, Z_3, \ldots\}$ is.
By Theorem~\ref{thm:concave}, we know that $E_r(t)$ is concave at $2, 3, \ldots$.
It is not hard to see that $\lim_{q \to \infty} \zeta_j^{i,r} = \delta_{j,\min\{i,r\}}$, where $\delta_{\cdot,\cdot}$ is the Kronecker delta.
We can then evaluate from \eqref{eq:ert} that for $r > 0$, $E_r(1) = 1$ and $E_r(2) = 1 + (1-p)$.
Then for $r > 0$, we have $1-p = E_r(2) - E_r(1) \le E_r(1) - E_r(0) = 1$, i.e., $E_r(t)$ is concave at $1$.
We do not need to check for $r = 0$ because $E_0(t)$ is a zero function which is always concave.
That is, the above non-stationary process gives a concave $E_r(t)$.

The concavity does not hold if we set $Z_t = 0$ in the above example.
In a similar fashion, we can verify that for $r > 0$, $E_r(1) = 0$ and $E_r(2) = 1-p$, thus $1-p = E_r(2) - E_r(1) \not\le E_r(1) - E_r(0) = 0$.
Loosely speaking, the concavity may hold if the channel gets ``worse'' over time, as the gain by sending one more packet is dropping over time.
We can see that this is the case in the above example with $Z_1 = 1$, i.e., the first packet is guaranteed to be received (and increase a rank) at the next node.
The channel becomes ``worse'' after that as there is a chance to lose a packet.
On the other hand, in the example with $Z_1 = 0$, i.e., the first packet must be lost, the channel becomes ``better'' after that as there is a chance to receive a packet (and increase a rank) at the next node.

\subsection{Properties of the Expected Rank Functions across Ranks}

We use the term $E_r(t+1) - E_r(t)$ very often in the remaining text, so we recall the notation in \eqref{eq:Delta} that $\Delta_{r,t} = E_r(t+1) - E_r(t)$.
We are going to investigate $\Delta_{r,t}$ with respect to $r$, which will be applied in Section~\ref{sec:approx}.

\begin{theorem} \label{thm:deltar}
	$\Delta_{r+1,t} \ge \Delta_{r,t}$.
	The inequality is strict if
	\begin{enumerate}[i)]
		\item $q$ is finite and $\Pr(Z_{t+1} = 1) \neq 0$; or
		\item $q \to \infty$ and $\Pr(Z_{t+1} = 1, \rk(\mathbf{Z}_t) = r) \neq 0$.
	\end{enumerate}
\end{theorem}

\begin{IEEEproof}
	See Appendix~\ref{sec:thm:deltar}.
\end{IEEEproof}

\begin{corollary} \label{cor:ratio}
	If $\{Z_t\}$ is a stationary stochastic process and $q \to \infty$, then $\Delta_{r,t} = \Delta_{r',t'}$, i.e., $E_r(t) : E_{r'}(t') = t : t'$, for all $t < r$ and $t' < r'$.
\end{corollary}

\begin{IEEEproof}
	By Theorem~\ref{thm:deltar}, we need $\Pr(Z_{t+1} = 1, \rk(\mathbf{Z}_t) = r) \neq 0$ to achieve $\Delta_{r+1,t} > \Delta_{r,t}$.
	However, this condition is not possible when $t < r$.
	Together with the linearity of $E_r(t)$ when $t < r$ by Corollary~\ref{cor:linear}, we have $\Delta_{r,t} = \Delta_{r',t'}$ for all $t < r$ and $t' < r'$.
	That is, we have $E_r(t) : E_{r'}(t') = t : t'$ for all $t < r$ and $t' < r'$.
\end{IEEEproof}

\begin{lemma} \label{lem:delta=}
	$\Delta\mathbb{E}[\rk(\mathbf{R}_{r+1,t+1})] = \Delta\mathbb{E}[\rk(\mathbf{R}_{r,t})]$ for all $r,t$ when $q \to \infty$.
\end{lemma}

\begin{IEEEproof}
	When $q \to \infty$, we have 
	$\rk(\mathbf{R}_{r,t}) = \min\{r,t\}$ with probability tends to $1$, which implies that $\mathbb{E}[\rk(\mathbf{R}_{r,t})] = \min\{r,t\}$.
	So, we have
	$\Delta\mathbb{E}[\rk(\mathbf{R}_{r,t})] = \Delta\mathbb{E}[\rk(\mathbf{R}_{r+1,t+1})]$ which equals $1$ if $t < r$; and $0$ otherwise.
\end{IEEEproof}

The numerical order between $\Delta\mathbb{E}[\rk(\mathbf{R}_{r,t})]$ and $\Delta\mathbb{E}[\rk(\mathbf{R}_{r+1,t+1})]$ will be used in Section~\ref{sec:approx} to show an intuition that the recoding number of a batch of a larger rank should be no smaller than that of a batch of a lower rank.
However for an arbitrary finite $q$, it is not straightforward to conclude such numerical order.
This is why we give a complicated assumption in the following theorem.
We defer the discussion about this to Appendix~\ref{sec:deltart}. %

\begin{theorem} \label{thm:diag}
	If $\{Z_t\}$ is a stationary stochastic process and
	\ifnum\paperversion=1
	\begin{multline}
		\sum_{i = 0}^t \Pr( Z_1 = 1, Z_{t+2} = 1, \rk(\mathbf{Y}_{t+1}) = i)\\
			\times (\Delta \mathbb{E}[\rk(\mathbf{R}_{r+1,i+1})] - \Delta \mathbb{E}[\rk(\mathbf{R}_{r,i})]) \ge 0, \label{eq:diag0}
	\end{multline}
	\else
	\begin{equation}
		\sum_{i = 0}^t \Pr( Z_1 = 1, Z_{t+2} = 1, \rk(\mathbf{Y}_{t+1}) = i) (\Delta \mathbb{E}[\rk(\mathbf{R}_{r+1,i+1})] - \Delta \mathbb{E}[\rk(\mathbf{R}_{r,i})]) \ge 0, \label{eq:diag0}
	\end{equation}
	\fi
	then $\Delta_{r+1,t+1} \ge \Delta_{r,t}$.
	The inequality is strict if
	\begin{enumerate}[i)]
		\item $q$ is finite and $\Pr(Z_1 = 0, Z_{t+2} = 1) \neq 0$; or
		\item $q \to \infty$ and $\Pr(Z_1 = 0, Z_{t+2} = 1, \rk(\mathbf{Y}_{t+1}) = r) \neq 0$.
	\end{enumerate}
\end{theorem}

\begin{IEEEproof}
	See Appendix~\ref{sec:thm:diag}.
\end{IEEEproof}

When $q \to \infty$, we can apply Lemma~\ref{lem:delta=} to show that the assumption \eqref{eq:diag0} in Theorem~\ref{thm:diag} holds.
However, the condition $\Pr(Z_1 = 0, Z_{t+2} = 1, \rk(\mathbf{Y}_{t+1}) = r) \neq 0$ for $\Delta_{r+1,t+1} > \Delta_{r,t}$ is not possible when $t < r$.

\section{Adaptive Recoding with Concave Expected Rank Functions} \label{sec:uni}

Recall the recoding number optimization problem \eqref{eq:P} and the optimization \eqref{eq:IP} formulated in Section~\ref{sec:adp}.
In this section, we characterize the solution of \eqref{eq:P} when the expected rank functions are concave.
We first show that the solution of \eqref{eq:IP} can also solve \eqref{eq:P}.
Then, we discuss further properties of the solution, such as the error tolerating property and the intuition that the number of recoded packets transmitted by a batch of a larger rank should be no less than the one of a batch of a smaller rank.

\subsection{Randomness on the Number of Recoded Packets}
\label{sec:ip}

Recall in \eqref{eq:realE} that we extend the domain of $E_r(t)$ from non-negative integers to non-negative real numbers by linear interpolation
	$E_r(t) = (t-\lfloor t \rfloor) E_r(\lfloor t \rfloor + 1) + (1 - (t-\lfloor t \rfloor)) E_r(\lfloor t \rfloor)$.
The following theorem suggests that we can solve \eqref{eq:IP} in lieu of \eqref{eq:P}.
As non-negative weighted sum of concave functions is concave \cite[p.~36]{pricing}, \eqref{eq:IP} is a concave optimization problem.
The implication is that for every rank $r$, the size of the support of $\alpha_{t|r}$ decreases to at most $2$.

\begin{theorem} \label{thm:IP}
	If $E_r(t)$ is concave for all $r \in [M]$, then an optimal solution of 
	\eqref{eq:IP}
	is an optimal solution of \eqref{eq:P}, where
	\begin{equation} \label{eq:IPalpha}
		\alpha_{t|r} = \begin{cases}
			t_r-\lfloor t_r \rfloor & \text{if } t = \lfloor t_r \rfloor + 1,\\
			1 - (t_r - \lfloor t_r \rfloor) & \text{if } t = \lfloor t_r \rfloor,\\
			0 & \text{otherwise.}
		\end{cases}
	\end{equation}
\end{theorem}

\begin{IEEEproof}
	Let $\{\alpha'_{t|r}\}$ be a feasible solution of \eqref{eq:P}, and let $t_r := \sum_{t = 0}^\infty t \alpha'_{t|r}$ be the mean of $\{\alpha'_{t|r}\}$.
	From the constraint $\sum_{r = 0}^M h_r \sum_{t = 0}^\infty t \alpha'_{t|r} = t_\text{avg}$ in \eqref{eq:P}, we know that $t_r$ converges.
	Also, $E_r(t)$ is defined and finite for all $t \ge 0$.
	As $E_r(t)$ is concave, the Jensen's inequality gives
	\begin{equation*}
		\sum_{t = 0}^\infty \alpha'_{t|r} E_r(t) \le E_r \left( \sum_{t = 0}^\infty t \alpha'_{t|r} \right) = E_r(t_r) = (1-(t_r-\lfloor t_r \rfloor)) E_r(\lfloor t_r \rfloor) + (t_r-\lfloor t_r \rfloor) E_r(\lfloor t_r \rfloor + 1).
	\end{equation*}
	Note that $\{\alpha_{t|r}\}$ is also a feasible solution of \eqref{eq:P}.
	As we can always achieve a non-decreased expected rank by using the mean $t_r$ of any feasible solution, there exists an optimal solution of \eqref{eq:P} in the form of \eqref{eq:IPalpha}.
	That is, an optimal solution of \eqref{eq:IP} is also an optimal solution of \eqref{eq:P}.
\end{IEEEproof}

We can interpret \eqref{eq:IP} as a network utility maximization problem \cite{NUM} by considering every rank as a user.
Suppose there are $M$ users.
The concave utility function for the user corresponds to the rank $r$ is $h_rE_r(t_r)$.
The \emph{resource} used by the users is the weighted sum $\sum_{r = 0}^M h_r t_r$, which cannot excess the available total resource $t_\text{avg}$.

Similar mathematical formulation also appears in other problems like the rate control in computer networks \cite{kelly,low} and the distributed control of electric vehicle charging \cite{ev}. %
These models consider multiple users in a general network and each user corresponds to one variable.
However in our case, all variables correspond to a single user for a single hop.

Up to this point, the worst case on the randomness in \eqref{eq:IP} is the case that all the $t_r$ are non-integers.
The following theorem states that there exists an optimal solution solving \eqref{eq:IP} such that there is at most one non-integer $t_r$, i.e., an almost deterministic solution (ADS), which minimize the randomness on the number of recoded packets.
This theorem can be proved by showing a non-decreased objective after eliminating the fractional parts of certain non-integer $t_r$ by reallocating the resource.

\begin{theorem} \label{thm:integer}
	There exists an optimal solution solving \eqref{eq:IP} such that there is at most one non-integer $t_r$.
\end{theorem}

\begin{IEEEproof}
	See Appendix~\ref{sec:thm:integer}.
\end{IEEEproof}

\subsection{Dual Formulation}

There is only one constraint in \eqref{eq:IP}, which means that its dual problem only has one Lagrange multiplier.
By Slater's condition, the strong duality holds for \eqref{eq:IP}.
The Lagrangian function is 
	$L(\{t_r\}_{r = 0}^M, \lambda) = \sum_{r = 0}^M h_r E_r(t_r) - \lambda \left( \sum_{r = 0}^M h_r t_r - t_\text{avg} \right)$,
where $\lambda \ge 0$ is the Lagrange multiplier.
The dual formulation of the problem \eqref{eq:IP} is
\begin{equation}
	\tag{D} \label{eq:D}
	\min_{\lambda \ge 0} \sup_{t_r \ge 0, \forall r \in [M]} L(\{t_r\}_{r = 0}^M, \lambda).
\end{equation}

The supremum of $L(\{t_r\}_{r = 0}^M, \lambda)$ can be rearranged into
	$\sum_{r = 0}^M \sup_{t_r \ge 0} ( h_r ( E_r(t_r) - \lambda t_r ) ) + \lambda t_\text{avg}$
so that we can apply dual decomposition \cite{dual_decomposition} on it.
That is, for each $r \in [M]$, we need to solve
\begin{equation} \label{eq:sup}
	\sup_{t_r \ge 0} ( h_r ( E_r(t_r) - \lambda t_r ) ).
\end{equation}

Recall that in \eqref{eq:Delta}, we defined $\Delta_{r,t} := E_r(t+1) - E_r(t)$.
On the other hand, $E_r(t)$ is formed by appending line segments of slopes $\Delta_{r,0}, \Delta_{r,1}, \ldots$ into a continuous curve, where $\Delta_{r,t}$ is non-increasing with respect to $t$.
So, the curve $E_r(t)-\lambda t_r$ is formed by appending line segments of slopes $\Delta_{r,0}-\lambda, \Delta_{r,1}-\lambda, \ldots$.
The optimal point is laid on the line segment before the slope becomes negative, or on some previous line segments if the slopes of the last few line segments are the same.
Let $E'_r(t)$ be the superdifferential of $E_r(t)$ at $t$, which is the interval of all the positive supergradients at $t$.
That is,
\begin{equation*}
	E'_r(t) = \begin{cases}
		[\Delta_{r,0}, \infty) & \text{if } t = 0,\\
		[\Delta_{r,t}, \Delta_{r,t-1}] & \text{if } t \in \mathbb{Z}^+,\\
		[\Delta_{r,t}, \Delta_{r,t}] & \text{otherwise}.
	\end{cases}
\end{equation*}

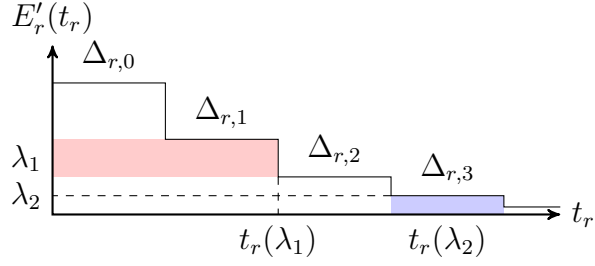
\begin{figure}
	\centering
	\begin{tikzpicture}[yscale=.5,xscale=1.5]
		\fill[red!20] (0,2) rectangle (2,1);
		\fill[blue!20] (3,.5) rectangle (4,0);
		\draw (0,3.5) -- (1,3.5) -- (1,2) -- (2,2) -- (2,1) -- (3,1) -- (3,.5) -- (4,.5) -- (4,.2) -- (4.5,.2);
		\node[left] at (0,1.5) {$\lambda_1$};
		\draw[dashed] (2,1) -- (2,0) node[below] {$t_r(\lambda_1)$};
		\draw[dashed] (3,.5) -- (0,.5) node[left] {$\lambda_2$};
		\node[below] at (3.5,0) {$t_r(\lambda_2)$};
		\draw [<->,thick] (0,4.5) node (yaxis) [above] {$E'_r(t_r)$} |- (4.5,0) node (xaxis) [right] {$t_r$};
		\node[above] at (.5,3.5) {$\Delta_{r,0}$};
		\node[above] at (1.5,2) {$\Delta_{r,1}$};
		\node[above] at (2.5,1) {$\Delta_{r,2}$};
		\node[above] at (3.5,.5) {$\Delta_{r,3}$};
	\end{tikzpicture}
	\caption{This figure illustrates the relationship between $\lambda$ and $t_r(\lambda)$, which is shown in \eqref{eq:tr}.
		The vertical lines only appear at integer $t_r$, which are the %
		superdifferential of $E_r(t_r)$ at $t_r$.
		All $\lambda_1$ in the red region share the same $t_r(\lambda_1)$, and all $t_r(\lambda_2)$ in the blue region share the same $\lambda_2$.
	}
	\label{fig:E'}
\end{figure}

With this notion, we can see that the $t_r$ solving \eqref{eq:sup} satisfies $\lambda \in E'_r(t)$.
If we write $t_r$ as a function of $\lambda$, then we have
\begin{equation} \label{eq:tr}
	\ifnum\paperversion=1
	\Scale[0.95]{
	\fi
	t_r(\lambda) = \begin{cases}
		[t+1,t+1] & \text{if } \Delta_{r,t} > \lambda > \Delta_{r,t+1},\\
		[\min\{a \colon \Delta_{r,a} = \lambda\}, t+1] & \text{if } \Delta_{r,t} = \lambda > \Delta_{r,t+1},\\
		[0,0] & \text{if } \lambda > \Delta_{r,0},
	\end{cases}
	\ifnum\paperversion=1
	}
	\fi
\end{equation}
where each $t_r(\lambda)$ is an interval.
Fig.~\ref{fig:E'} illustrates the relationship between $\lambda$ and $t_r(\lambda)$.
In other words, the Lagrange multiplier indicates the rate of change of $E_r(t)$, i.e., $E'_r(t)$, for all $r \in [M]$ at the same time, where each of which corresponds to an interval for the primal variable.

\subsection{A $3$-D Illustration}
\label{sec:3d}

\begin{figure}
	\centering
	\begin{minipage}[c]{.5\textwidth}
		\includegraphics[scale=.5]{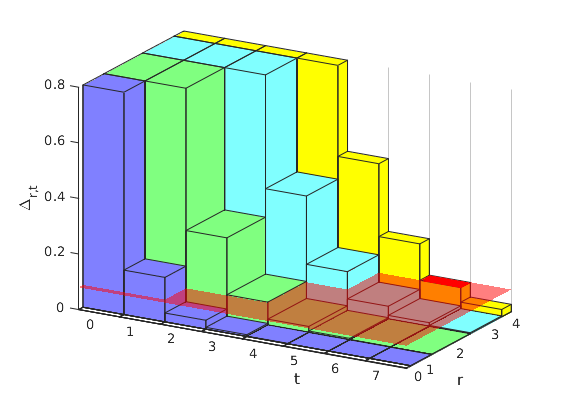}
		\caption{A $3$-D illustration of the whole picture.}
		\label{fig:3dplot}
	\end{minipage}~
	\begin{minipage}[c]{.5\textwidth}
		\centering
		\includegraphics[scale=.5]{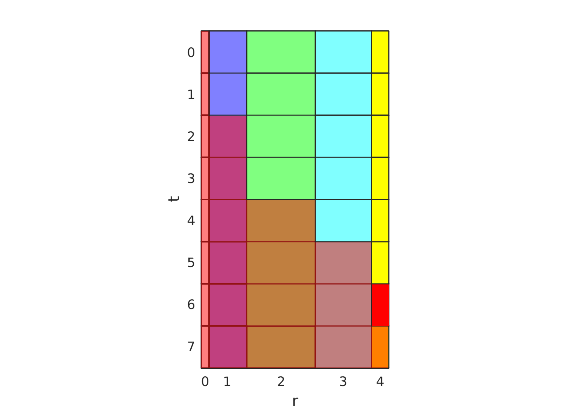}
		\caption{The top view of Fig.~\ref{fig:3dplot}.}
		\label{fig:topview}
	\end{minipage}
\end{figure}

We can see the relation between \eqref{eq:IP}, \eqref{eq:D}, the allocated resource and the expected rank functions all together in Fig.~\ref{fig:3dplot}.
In the figure, the width for each $r$ in the $r$-axis is $h_r$, while the width for each $t$ is $1$.
The height of the bar located at $(r, t)$ is $\Delta_{r,t}$, which is monotonic decreasing when $t$ increases.
The red semitransparent surface acts as a ``water surface'', where its height is the Lagrange multiplier.

The cross-section of Fig.~\ref{fig:3dplot} by fixing an $r$ is the one illustrated in Fig.~\ref{fig:E'} (after shifting the $t$-axis to the left so that the top of each bar lays between two integers).
That is, in the view of Fig.~\ref{fig:3dplot}, $t_r(\lambda)$ corresponds to the shortest bar of the corresponding $r$ which is above the water surface.
If the water surface touches the top of the shortest bar, e.g., the solid red top face located at $(4, 6)$, then the corresponding $t_r$ can be a non-integer.

We can illustrate the allocated resource from Fig.~\ref{fig:topview}, which is the top view of Fig.~\ref{fig:3dplot}.
For each $r$, the sum of the areas of the top faces, i.e., the sum of the areas of the rectangles of the same color strip which are not covered in red in Fig.~\ref{fig:topview}, is $h_rt_r$.
The solid red face is an exception in a way that only a portion of the area is counted as this face corresponds to a non-integer $t_r$.
In other words, the sum of the areas of the rectangles which are not covered in red in Fig.~\ref{fig:topview}, plus a portion of the area of the solid red face, is the allocated resource.

Note that from the definition of $\Delta_{r,t}$, we can express $E_r(t_r)$ by
\begin{equation} \label{eq:edelta}
	E_r(t_r) = \sum_{i = 0}^{\lfloor t_r \rfloor - 1} \Delta_{r,i} + (t_r - \lfloor t_r \rfloor) \Delta_{r, \lfloor t_r \rfloor}.
\end{equation}
Therefore, the objective value of \eqref{eq:IP}, $\sum_{r = 0}^M h_r E_r(t_r)$, is the sum of the volumes of the bars contributed to the areas for allocated resource, i.e., the volume of those bars which are taller than the water surface.
Again, only a portion of the volume of bar with the solid red top face is counted.

From Fig.~\ref{fig:3dplot}, we can also explain why our Algorithm~\ref{alg:opt} works.
Suppose we start with a certain height of the water surface.
The bars above the water surface gives an interior point as the starting point of the algorithm.
We can set a water surface higher than the tallest bar so that we start from $t_r = 0$ for all $r \in [M]$.
We lower the water surface in each iteration until a new top face comes out from the water.
This top face belongs to the tallest bar compared with those bars under the water.
The height of this bar equals the largest $\Delta_{r,t_r}$ at this moment.
When we lower the water surface to a point where the allocated resource (the area viewed from the top view) equals the total resource, then the algorithm stops.
A special handling of the solid red face is done to calculate the portion of its area and volume to be included in the allocated resource and the objective respectively.
This way, we can ensure that only those tallest bars are out of the water surface, thus the sum of volumes is maximized.

\subsection{Multiset Representation}
\label{sec:2f}

To further investigate the characteristic of the solution, e.g., the error toleration on the input rank distribution, and the behavior of the algorithms, e.g., starting from an arbitrary interior point which does not follow the water surface criteria, we present a useful representation of an optimal solution of \eqref{eq:IP} and also an arbitrary interior point.

For any $r \in [M]$, define a function $\Omega_r$ which maps a non-negative real number to a multiset:
\begin{equation*}
	\Omega_r(t_r) := \{\Delta_{r,t} \colon t < t_r, t \in \mathbb{N}\}.
\end{equation*}
For simplicity, define $\Omega_r(\infty) := \{\Delta_{r,t} \colon t \in \mathbb{N}\}$.
We use the notation $\uplus$ to represent the additive union of multisets \cite{multiset}.

Due to the property that $\Delta_{r,t} \ge \Delta_{r,t+1}$, every element in $\Omega_r(\infty) \setminus \Omega_r(t_r)$ is no larger than the elements in $\Omega_r(t_r)$ when $\Omega_r(t_r) \neq \emptyset$.
That is, we have $\min \Omega_r(t_r) = \Delta_{r, \lceil t_r-1 \rceil}$ when $t_r > 0$.
Also, the largest $\lceil t_r \rceil$ elements in $\Omega_r(\infty)$, which is exactly $\Omega_r(t_r)$, can be used to represent the terms involved in \eqref{eq:edelta}.
These properties can be easily seen in the $3$-D illustration in Fig.~\ref{fig:3dplot}.

\begin{figure}
	\scriptsize
	\centering
	\begin{tikzpicture}[scale=.5]
		\fill[red!20] (5,0) rectangle ++(1,4);
		\node at (5.4,4.3) {$E_5(8)$};
		\fill[red!20] (3,0) rectangle ++(1,2.25);
		\node at (3.4,2.7) {$E_3(4.5)$};
		\draw[xscale=1,yscale=.5,step=1,gray,very thin] (0,0) grid (8,10);
		\draw[<->,thick] (0,5) node (yaxis) [above] {$t$} |- (8,0) node (xaxis) [right] {$r$};
		\foreach \i in {0,...,7}
		{
			\node at (.5+\i,-.3) {$\i$};
		}
		\foreach \j in {0,...,9}
		{
			\node at (-.2,.25+.5*\j) {$\j$};
		}
	\end{tikzpicture}
	\caption{The representation of expected rank functions in tabular form.}
	\label{fig:tabular}
\end{figure}
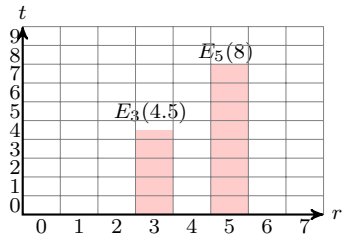

Now, we use a table of cells as shown in Fig.~\ref{fig:tabular} to illustrate the above mathematical structures.
The table reassembles the top view in Fig.~\ref{fig:topview} by neglecting the widths of the bars in the $r$-axis.
This way, we can also visualize those $r$ where $h_r = 0$.
Each cell in Fig.~\ref{fig:tabular} corresponds to a value of $\Delta_{r,t}$.
The lower $\lceil t_r \rceil$ cells in column $r$ are all the elements in $\Omega_r(t_r)$.
In other words, $\Omega_r(\infty)$ can be represented by all the cells in column $r$.

We can use this relation to express $E_r(t_r)$ in the table.
For example, $E_5(8)$ is the sum of the lower $8$ cells in column $5$.
When $t_r$ is not an integer, we can represent the fractional part by taking a part of a cell.
For example, $E_3(4.5)$ is the sum of the lower $4$ cells plus half of the $5$-th cell in column $3$.
The lower $\lceil 4.5 \rceil = 5$ cells in the column are involved in expressing $E_3(4.5)$.

If there is some $r$ where $h_r = 0$, an arbitrary $t_r$ would not consume any resource nor affect the objective value in \eqref{eq:IP}.
If the zero mass is due to certain error in measurement, we may receive a batch of rank $r$ and we need to make a reasonable decision on its recoding number.
We defer the discussion of this case to the next subsection.
In the remainder of this subsection, we concern those $r$ where $h_r \neq 0$.

\begin{definition}[Rank Support] \label{def:ranksupport}
	The set $S := \{r \colon h_r \neq 0\}$ is called the \emph{rank support}.
\end{definition}

\begin{theorem} \label{thm:IPomega}
	If $\{t_r\}_{r = 0}^M$ is an optimal solution of \eqref{eq:IP}, then $\biguplus_{r \in S} \Omega_r(t_r)$ is a collection of the largest $\sum_{r \in S} \lceil t_r \rceil$ elements in $\biguplus_{r \in S} \Omega_r(\infty)$.
\end{theorem}

\begin{IEEEproof}
	See Appendix~\ref{sec:thm:IPomega}.
\end{IEEEproof}

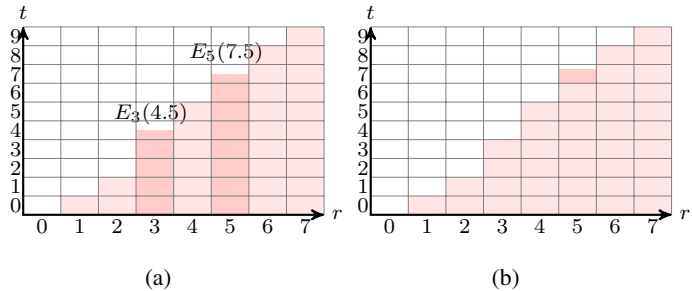
\begin{figure}
	\scriptsize
	\centering
	\begin{subfigure}{.25\textwidth}
		\centering
		\begin{tikzpicture}[scale=.5] %
			\fill[red!10] (1,0) rectangle ++(1,.5);
			\fill[red!10] (2,0) rectangle ++(1,1);
			\fill[red!10] (4,0) rectangle ++(1,3);
			\fill[red!10] (6,0) rectangle ++(1,4.5);
			\fill[red!10] (7,0) rectangle ++(1,5);
			\fill[red!20] (5,0) rectangle ++(1,3.75);
			\node at (5.4,4.3) {$E_5(7.5)$};
			\fill[red!20] (3,0) rectangle ++(1,2.25);
			\node at (3.4,2.7) {$E_3(4.5)$};
			\draw[xscale=1,yscale=.5,step=1,gray,very thin] (0,0) grid (8,10);
			\draw[<->,thick] (0,5) node (yaxis) [above] {$t$} |- (8,0) node (xaxis) [right] {$r$};
			\foreach \i in {0,...,7}
			{
				\node at (.5+\i,-.3) {$\i$};
			}
			\foreach \j in {0,...,9}
			{
				\node at (-.2,.25+.5*\j) {$\j$};
			}
		\end{tikzpicture}
		\caption{}
		\label{fig:tab_opt1}
	\end{subfigure}~~~~
	\begin{subfigure}{.25\textwidth}
		\centering
		\begin{tikzpicture}[scale=.5] %
			\fill[red!10] (1,0) rectangle ++(1,.5);
			\fill[red!10] (2,0) rectangle ++(1,1);
			\fill[red!10] (4,0) rectangle ++(1,3);
			\fill[red!10] (6,0) rectangle ++(1,4.5);
			\fill[red!10] (7,0) rectangle ++(1,5);
			\fill[red!10] (5,0) rectangle ++(1,3.5);
			\fill[red!20] (5,3.5) rectangle ++(1,.375);
			\fill[red!10] (3,0) rectangle ++(1,2);
			\draw[xscale=1,yscale=.5,step=1,gray,very thin] (0,0) grid (8,10);
			\draw[<->,thick] (0,5) node (yaxis) [above] {$t$} |- (8,0) node (xaxis) [right] {$r$};
			\foreach \i in {0,...,7}
			{
				\node at (.5+\i,-.3) {$\i$};
			}
			\foreach \j in {0,...,9}
			{
				\node at (-.2,.25+.5*\j) {$\j$};
			}
		\end{tikzpicture}
		\caption{}
		\label{fig:tab_opt2}
	\end{subfigure}
	\caption{Illustration of some feasible solutions.}
\end{figure}

Theorem~\ref{thm:IPomega} does not restrict the number of non-integer $t_r$.
For example, we may have an optimal solution as illustrated in Fig.~\ref{fig:tab_opt1}.
The two partially filled cells must have the same value, or otherwise we can reallocate the resource consumed by these two cells to obtain a larger objective.
We can reallocate the resource consumed by these two cells to obtain another optimal solution with at most one non-integer $t_r$.
One possibility is illustrated in Fig.~\ref{fig:tab_opt2}.

The converse of Theorem~\ref{thm:IPomega} is not always true.
For example, suppose we have a feasible solution $\{t_r\}_{r = 0}^M$ and distinct $m, n \in S$ such that $t_m, t_n$ are non-integers and $\Delta_{m,\lfloor t_m \rfloor} > \Delta_{n, \lfloor t_n \rfloor}$, which is illustrated in Fig.~\ref{fig:tab_opt1}.
It is possible that $\biguplus_{r \in S} \Omega_r(t_r)$ is a collection of the largest $\sum_{r \in S} \lceil t_r \rceil$ elements in $\biguplus_{r \in S} \Omega_r(\infty)$, but it is clear that $\{t_r\}_{r = 0}^M$ is not an optimal solution as we can reallocate the resource to increase $t_m$ and decrease $t_n$ by a little bit which yields a higher objective value.
However, the following theorem states a special case where the converse is true.

\begin{theorem} \label{thm:omegaIP}
	Let $\{t_r\}_{r = 0}^M$ be a feasible solution.
	If
	\begin{enumerate}[i)]
		\item there is at most one non-integer in $\{t_r\}_{r \in S}$, and the non-integer $t_r$ can only appear on an $r$ where $\Delta_{r,\lceil t_r-1 \rceil} = \min \biguplus_{r \in S} \Omega_r(t_r)$; and
		\item $\biguplus_{r \in S} \Omega_r(t_r)$ is a collection of the largest $\sum_{r \in S} \lceil t_r \rceil$ elements in $\biguplus_{r \in S} \Omega_r(\infty)$,
	\end{enumerate}
	then $\{t_r\}_{r = 0}^M$ is an optimal solution of \eqref{eq:IP}.
\end{theorem}

\begin{IEEEproof}
	See Appendix~\ref{sec:thm:omegaIP}.
\end{IEEEproof}

\begin{corollary} \label{cor:tavg}
	There exists some $t_\text{avg} \ge 0$ such that $\{t_r\}_{r = 0}^M$ is an optimal solution of \eqref{eq:IP} where $t_r$ are all integers.
\end{corollary}

\begin{IEEEproof}
	The case $t_\text{avg} = 0$ is trivial.
	For $t_\text{avg} > 0$, we select a non-zero amount of largest elements in $\biguplus_{r \in S} \Omega_r(\infty)$.
	Then, the collection can be expressed by $\biguplus_{r \in S} \Omega_r(t_r)$ for some integers $t_r$.
	We can choose an arbitrary integer $t_r$ for those $r \not \in S$.
	To become a feasible solution, we have to select $t_\text{avg} = \sum_{r = 0}^M h_r t_r > 0$.
	By Theorem~\ref{thm:omegaIP}, the proof is done.
\end{IEEEproof}

Corollary~\ref{cor:tavg} suggests that if we are allowed to choose the amount of total resource $t_\text{avg}$, there exists a choice such that the optimal recoding numbers are deterministic.

\subsection{Error Toleration on the Input Rank Distribution} \label{sec:preferred}

Note that when we express an optimal solution $\{t_r\}_{r = 0}^M$ in the form of the largest $\sum_{r \in S} \lceil t_r \rceil$ elements in $\biguplus_{r \in S} \Omega_r(\infty)$, the value of the input rank distribution $(h_0, \ldots, h_M)$ is embedded in the numerical value of $t_r$.
In other words, the input rank distribution only has impact on the number of largest elements in the collection we have selected, which is constrained by the total resource $\sum_{r = 0}^M h_r t_r = t_\text{avg}$.

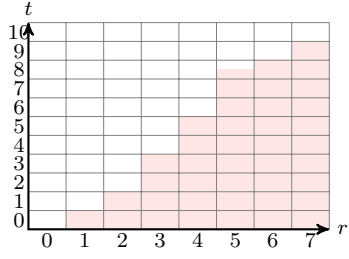
\begin{figure}
	\scriptsize
	\centering
	\begin{tikzpicture}[scale=.5]
		\fill[red!10] (1,0) rectangle ++(1,.5);
		\fill[red!10] (2,0) rectangle ++(1,1);
		\fill[red!10] (4,0) rectangle ++(1,3);
		\fill[red!10] (6,0) rectangle ++(1,4.5);
		\fill[red!10] (7,0) rectangle ++(1,5);
		\fill[red!10] (5,0) rectangle ++(1,4);
		\fill[red!10] (5,4) rectangle ++(1,.25);
		\fill[red!10] (3,0) rectangle ++(1,2);
		\draw[xscale=1,yscale=.5,step=1,gray,very thin] (0,0) grid (8,11);
		\draw[<->,thick] (0,5.5) node (yaxis) [above] {$t$} |- (8,0) node (xaxis) [right] {$r$};
		\foreach \i in {0,...,7}
		{
			\node at (.5+\i,-.3) {$\i$};
		}
		\foreach \j in {0,...,10}
		{
			\node at (-.25,.25+.5*\j) {$\j$};
		}
	\end{tikzpicture}
	\caption{An example of optimal solution with an accurate input rank distribution.}
	\label{fig:tab_opt3}
\end{figure}

We consider a solution satisfying the conditions in Theorem~\ref{thm:omegaIP}.
Take Fig.~\ref{fig:tab_opt3} as an example of such a solution.
Suppose we have some errors on some $h_r$ where $r \in S$.
Let $(h'_0, \ldots, h'_M)$ be the true input rank distribution.
\begin{itemize}
	\item Case I: $\sum_{r = 0}^M h'_r t_r = t_\text{avg}$.
We do not need to modify the selected collection.
	As the selected collection are the largest cells, we know that $\{t_r\}_{r = 0}^M$ is also an optimal solution when we use the true input rank distribution.

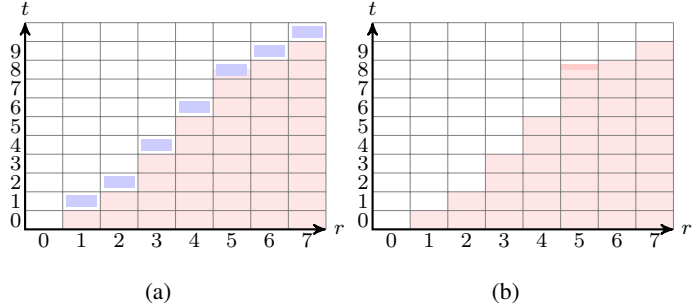
\begin{figure}
	\scriptsize
	\centering
	\begin{subfigure}{.25\textwidth}
		\centering
		\begin{tikzpicture}[scale=.5]
			\fill[red!10] (1,0) rectangle ++(1,.5);
			\fill[red!10] (2,0) rectangle ++(1,1);
			\fill[red!10] (4,0) rectangle ++(1,3);
			\fill[red!10] (6,0) rectangle ++(1,4.5);
			\fill[red!10] (7,0) rectangle ++(1,5);
			\fill[red!10] (5,0) rectangle ++(1,4);
			\fill[red!10] (5,4) rectangle ++(1,.25);
			\fill[red!10] (3,0) rectangle ++(1,2);
			\fill[blue!20] (1.1,.6) rectangle ++(.8,.3);
			\fill[blue!20] (2.1,1.1) rectangle ++(.8,.3);
			\fill[blue!20] (3.1,2.1) rectangle ++(.8,.3);
			\fill[blue!20] (4.1,3.1) rectangle ++(.8,.3);
			\fill[blue!20] (5.1,4.1) rectangle ++(.8,.3);
			\fill[blue!20] (6.1,4.6) rectangle ++(.8,.3);
			\fill[blue!20] (7.1,5.1) rectangle ++(.8,.3);
			\draw[xscale=1,yscale=.5,step=1,gray,very thin] (0,0) grid (8,11);
			\draw[<->,thick] (0,5.5) node (yaxis) [above] {$t$} |- (8,0) node (xaxis) [right] {$r$};
			\foreach \i in {0,...,7}
			{
				\node at (.5+\i,-.3) {$\i$};
			}
			\foreach \j in {0,...,9}
			{
				\node at (-.25,.25+.5*\j) {$\j$};
			}
		\end{tikzpicture}
		\caption{}
		\label{fig:tab_case2a}
	\end{subfigure}~~~~
	\begin{subfigure}{.25\textwidth}
		\centering
		\begin{tikzpicture}[scale=.5]
			\fill[red!10] (1,0) rectangle ++(1,.5);
			\fill[red!10] (2,0) rectangle ++(1,1);
			\fill[red!10] (4,0) rectangle ++(1,3);
			\fill[red!10] (6,0) rectangle ++(1,4.5);
			\fill[red!10] (7,0) rectangle ++(1,5);
			\fill[red!10] (5,0) rectangle ++(1,4);
			\fill[red!10] (5,4) rectangle ++(1,.25);
			\fill[red!10] (3,0) rectangle ++(1,2);
			\fill[red!20] (5,4.25) rectangle ++(1,.15);
			\draw[xscale=1,yscale=.5,step=1,gray,very thin] (0,0) grid (8,11);
			\draw[<->,thick] (0,5.5) node (yaxis) [above] {$t$} |- (8,0) node (xaxis) [right] {$r$};
			\foreach \i in {0,...,7}
			{
				\node at (.5+\i,-.3) {$\i$};
			}
			\foreach \j in {0,...,9}
			{
				\node at (-.25,.25+.5*\j) {$\j$};
			}
		\end{tikzpicture}
		\caption{}
		\label{fig:tab_case2b}
	\end{subfigure}
	\caption{Occupying more resource.}
\end{figure}

\item Case II: $\sum_{r = 0}^M h'_r t_r < t_\text{avg}$.
We need to allocate more resource.
That is, we have to include some largest elements in $\biguplus_{r \in S} (\Omega_r(\infty) \setminus \Omega_r(t_r))$ when necessary.
Again, this keeps the selected collection as the largest elements in $\biguplus_{r \in S} \Omega_r(\infty)$.
As an example, the blue cells in Fig.~\ref{fig:tab_case2a} are those largest elements.
We select the blue cell with the largest value and occupy more resource from it.
If the blue cells are completely filled, we need to find another set of blue cells.
Similarly to the previous case, when the change in resource is small enough, only one of the $t_r$ will be changed a little bit as shown in Fig.~\ref{fig:tab_case2b}.

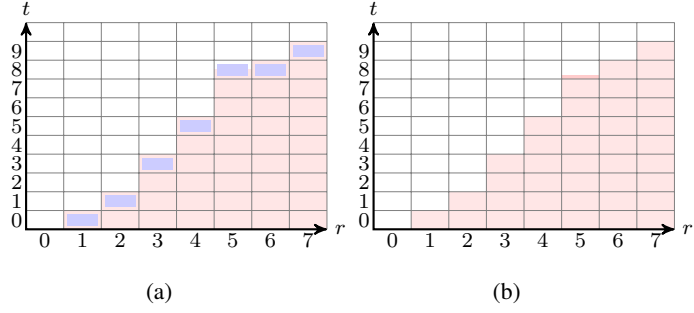
\begin{figure}
	\scriptsize
	\centering
	\begin{subfigure}{.25\textwidth}
		\centering
		\begin{tikzpicture}[scale=.5]
			\fill[red!10] (1,0) rectangle ++(1,.5);
			\fill[red!10] (2,0) rectangle ++(1,1);
			\fill[red!10] (4,0) rectangle ++(1,3);
			\fill[red!10] (6,0) rectangle ++(1,4.5);
			\fill[red!10] (7,0) rectangle ++(1,5);
			\fill[red!10] (5,0) rectangle ++(1,4);
			\fill[red!10] (5,4) rectangle ++(1,.25);
			\fill[red!10] (3,0) rectangle ++(1,2);
			\fill[blue!20] (1.1,.1) rectangle ++(.8,.3);
			\fill[blue!20] (2.1,.6) rectangle ++(.8,.3);
			\fill[blue!20] (3.1,1.6) rectangle ++(.8,.3);
			\fill[blue!20] (4.1,2.6) rectangle ++(.8,.3);
			\fill[blue!20] (5.1,4.1) rectangle ++(.8,.3);
			\fill[blue!20] (6.1,4.1) rectangle ++(.8,.3);
			\fill[blue!20] (7.1,4.6) rectangle ++(.8,.3);
			\draw[xscale=1,yscale=.5,step=1,gray,very thin] (0,0) grid (8,11);
			\draw[<->,thick] (0,5.5) node (yaxis) [above] {$t$} |- (8,0) node (xaxis) [right] {$r$};
			\foreach \i in {0,...,7}
			{
				\node at (.5+\i,-.3) {$\i$};
			}
			\foreach \j in {0,...,9}
			{
				\node at (-.25,.25+.5*\j) {$\j$};
			}
		\end{tikzpicture}
		\caption{}
		\label{fig:tab_case3a}
	\end{subfigure}~~~~
	\begin{subfigure}{.25\textwidth}
		\centering
		\begin{tikzpicture}[scale=.5]
			\fill[red!10] (1,0) rectangle ++(1,.5);
			\fill[red!10] (2,0) rectangle ++(1,1);
			\fill[red!10] (4,0) rectangle ++(1,3);
			\fill[red!10] (6,0) rectangle ++(1,4.5);
			\fill[red!10] (7,0) rectangle ++(1,5);
			\fill[red!10] (5,0) rectangle ++(1,4);
			\fill[red!10] (3,0) rectangle ++(1,2);
			\fill[red!20] (5,4) rectangle ++(1,.1);
			\draw[xscale=1,yscale=.5,step=1,gray,very thin] (0,0) grid (8,11);
			\draw[<->,thick] (0,5.5) node (yaxis) [above] {$t$} |- (8,0) node (xaxis) [right] {$r$};
			\foreach \i in {0,...,7}
			{
				\node at (.5+\i,-.3) {$\i$};
			}
			\foreach \j in {0,...,9}
			{
				\node at (-.25,.25+.5*\j) {$\j$};
			}
		\end{tikzpicture}
		\caption{}
		\label{fig:tab_case3b}
	\end{subfigure}
	\caption{Releasing some occupied resource.}
\end{figure}

\item Case III: $\sum_{r = 0}^M h'_r t_r > t_\text{avg}$.
We need to release some occupied resource.
It means that we are going to remove some smallest elements in $\biguplus_{r \in S} \Omega_r(t_r)$, if necessary.
This procedure keeps the largest elements selected.
As an example, the blue cells in Fig.~\ref{fig:tab_case3a} are those smallest elements.
We release the resource from one blue cell which has the smallest value.
If the blue cells are completely drained, we need to find another set of blue cells.
When the change in resource is small enough, only one of the $t_r$ will be changed a little bit as shown in Fig.~\ref{fig:tab_case3b}.
\end{itemize}

As a summary, when $|\sum_{r = 0}^M (h'_r - h_r) t_r|$ is small, the above discussion suggests that the change in $\{t_r\}_{r \in S}$ is not large.
That is, $\{t_r\}_{r \in S}$ is error tolerating on the input rank distribution.
However, it is not the case when $r \not \in S$.
If there is an error on $h_{r'}$ where $r' \not \in S$, then the corresponding $t_{r'}$ is no longer arbitrary.
We have to include $\Omega_{r'}(\infty)$ in the selection of the largest elements, which can have a huge change from the original $t_{r'}$.

As we want $\{t_r\}_{r \not \in S}$ to be error tolerating, an easy approach is to let those $r \not \in S$ having a similar behavior as those $r \in S$.
In light of this, we define the preferred solution as follows.

\begin{definition}[Preferred Solution]
	A feasible solution $\{t_r\}_{r = 0}^M$ of \eqref{eq:IP} is a \emph{preferred solution} if
\begin{enumerate}[i)]
	\item there is at most one non-integer $t_r$, and the non-integer $t_r$ can only appears on an $r$ where $\Delta_{r, \lceil t_r-1 \rceil} = \min \biguplus_{r \in [M]} \Omega_r(t_r)$; and
	\item $\biguplus_{r \in [M]} \Omega_r(t_r)$ is a collection of the largest $\sum_{r \in [M]} \lceil t_r \rceil$ elements in $\biguplus_{r \in [M]} \Omega_r(\infty)$.
\end{enumerate}
\end{definition}

By Theorem~\ref{thm:omegaIP}, a preferred solution is an optimal solution.
We can see that this definition implies the same meaning as in Definition~\ref{def:preferred}.

\subsection{Other Characteristics of Optimal Solutions}
\label{sec:approx}

Intuitively, we should transmit more recoded packets for a batch of a larger rank.
A weaker version of this intuition is that, the recoding number of a batch of a larger rank should be no smaller than that of a batch of a lower rank.
The following theorem gives a sufficient condition on the above intuitions.

\begin{theorem} \label{thm:intuition}
	If $\Delta_{r+1,t} > \Delta_{r,t}$ for all $r, t$, then the optimal solution of \eqref{eq:IP} satisfies $t_n \ge t_m$ for all $n > m$ such that $n, m \in S$ and $t_m > 0$.
	The inequality is strict if $\Delta_{r+1,t+1} > \Delta_{r,t}$ for all $r,t$.
Further, if the optimal solution $\{t_r\}_{r = 0}^M$ must satisfy $t_r \ge r$ for all $r \in S$, then $t_n > t_m$ if $\Delta_{r+1,t+1} > \Delta_{r,t}$ for all $t \ge r$.
\end{theorem}

\begin{IEEEproof}
	See Appendix~\ref{sec:thm:intuition}.
\end{IEEEproof}

This intuition inspired different approximation schemes for adaptive recoding.
For example, the one proposed in \cite{adaptive} for independent packet loss channel first assigns $t_r = r + \gamma$ for all $r \in [M]$ where $\gamma \in \mathbb{N}$, and then assigns the remaining resource to those with the highest ranks.
Another example is the one proposed in \cite{ge_adaptive} for GE model which assigns $t_r \propto r$ for all $r \in [M]$ and performs certain rounding afterward.
Both works assume $q \to \infty$.
To see that the intuition is true in these scenarios, we consider the following corollary.
Recall %
that $\mathbf{Y}_t := \diag(Z_2, \ldots, Z_t)$.

\begin{corollary} \label{cor:rhr}
	If $\{Z_t\}$ is a stationary stochastic process, $q \to \infty$ and $t_\text{avg} \ge \sum_{r = 0}^M r h_r$, then there is an optimal solution $\{t_r\}_{r = 0}^M$ of \eqref{eq:IP} such that $t_r \ge r$ for all $r \in [M]$.
	Further, if $\Pr(Z_1 = 1. Z_{c+1} = 1, \rk(\mathbf{Y}_c) = r-1) \neq 0$ for all $c \ge r$, then any optimal solution must satisfy $t_r \ge r$ for all $r \in S$.
\end{corollary}

\begin{IEEEproof}
	See Appendix~\ref{sec:cor:rhr}.
\end{IEEEproof}

The independent loss channel model and the GE model satisfy the following:
\begin{itemize}
	\item $\Pr(Z_1 = 1. Z_{t+1} = 1, \rk(\mathbf{Y}_t) = r-1) \neq 0$;
	\item $\Pr(Z_{t+1} = 1, \rk(\mathbf{Z}_t) = r) \neq 0$; and
	\item $\Pr(Z_1 = 0, Z_{t+2} = 1, \rk(\mathbf{Y}_{t+1}) = r) \neq 0$
\end{itemize}
for all $t \ge r$.
When we consider $q \to \infty$, we have
\begin{itemize}
	\item $t_r \ge r$ for all $r \in S$ by Corollary~\ref{cor:rhr};
	\item $\Delta_{r+1,t} > \Delta_{r,t}$ for all $t \ge r$ by Theorem~\ref{thm:deltar}; and
	\item $\Delta_{r+1,t+1} > \Delta_{r,t}$ for all $t \ge r$ by Theorem~\ref{thm:diag} and Lemma~\ref{lem:delta=}.
\end{itemize}
Then, the assumptions in Theorem~\ref{thm:intuition} are all satisfied, which show that the stronger intuition, i.e., a batch of a larger rank transmits strictly more recoded packets than a batch of lower rank, is valid.

When $q$ is finite, we still have $\Delta_{r+1,t} > \Delta_{r,t}$ for all $r,t$ by Theorem~\ref{thm:deltar}, so we can apply Theorem~\ref{thm:intuition} to show that we must have $t_n \ge t_m$ for all $n > m$ such that $n, m \in S$ and $t_m > 0$.
This means that even when we do not know the packet loss probability or the transition probabilities, the weaker intuition still holds.

\section{Algorithms} \label{sec:tune}

We want to find a preferred solution defined in Section~\ref{sec:preferred}, which can minimize the randomness on the number of recoded packets and tolerate rank distribution errors.
Although a preferred solution is an optimal solution, the converse is not necessary true.
Therefore, it is not guaranteed that the optimal solution given by an arbitrary optimization solver is a preferred solution.
We need tuning algorithms to produce a preferred solution.

\begin{figure}
\removelatexerror
\begin{algorithm}[H]
	\footnotesize
	\caption{Tuning from a feasible solution}
	\label{alg:tune}
	\KwData{A feasible solution $\{t_r\}_{r = 0}^M$, the input rank distribution $(h_0,\ldots,h_M)$, and the access to the values $\Delta_{r,t}$.}
	\KwResult{A preferred solution $\{t_r\}_{r = 0}^M$.}
	\While{$\max_{r \in [M]} \Delta_{r,\lfloor t_r \rfloor} > \min_{r \in [M]} \Delta_{r,\lceil t_r-1 \rceil}$}{
		$m \leftarrow$ an element in $\argmax_{r \in [M]} \Delta_{r,\lfloor t_r \rfloor}$ ;
		$n \leftarrow$ an element in $\argmin_{r \in [M]} \Delta_{r,\lceil t_r-1 \rceil}$ \;
		\If{$h_m = 0$}{
			$t_m \leftarrow \lfloor t_m + 1 \rfloor$ ;
			\Continue \;
		}
		\If{$h_n = 0$}{
			$t_n \leftarrow \lceil t_n - 1 \rceil$ ;
			\Continue \;
		}
		$s \leftarrow \min\{h_n(t_n - \lfloor t_n \rfloor + \delta_{\lfloor t_n \rfloor, t_n}), h_m (1-(t_m - \lfloor t_m \rfloor))\}$; $t_m \leftarrow t_m + s/h_m$; $t_n \leftarrow t_n - s/h_n$ \;
	}
	$u \leftarrow \sum_{r \in [M]} h_r (t_r - \lfloor t_r \rfloor)$ ;
	$t_r \leftarrow \lfloor t_r \rfloor, \forall r \in [M]$ \;
	\Return output of Algorithm~\ref{alg:opt} with input $\{t_r\}$ and $u$ \;
\end{algorithm}
\end{figure}

The tuning procedure includes two algorithms, which are the generalized versions of the algorithms in \cite{adaptive}.
The first algorithm is Algorithm~\ref{alg:opt}, which has been repeatedly mentioned throughout the paper.
The idea of the algorithm is that we occupy the largest cells one by one, so eventually we move an interior point to a feasible solution, but the optimality is not always guaranteed.
Under certain condition, Algorithm~\ref{alg:opt} can output a preferred solution.
The second algorithm, Algorithm~\ref{alg:tune}, tunes a feasible solution, regardless of its optimality, into a preferred solution.
The idea is to reallocate some resource to a point that satisfies the condition for Algorithm~\ref{alg:opt} to output a preferred solution.
In other words, Algorithm~\ref{alg:tune} internally calls Algorithm~\ref{alg:opt}.

\subsection{From an Interior Point to a Feasible Solution}
\label{sec:primal}

Algorithm~\ref{alg:opt} has been appeared throughout this paper.
In this subsection, we discuss the limitation and correctness of this algorithm.

\begin{figure}
	\scriptsize
	\centering
	\begin{tikzpicture}[scale=.5]
		\fill[red!10] (1,0) rectangle ++(1,.5);
		\fill[red!10] (2,0) rectangle ++(1,1);
		\fill[red!10] (4,0) rectangle ++(1,3);
		\fill[red!10] (6,0) rectangle ++(1,4.5);
		\fill[red!10] (7,0) rectangle ++(1,5);
		\fill[red!10] (5,0) rectangle ++(1,4.25);
		\fill[red!10] (3,0) rectangle ++(1,2);
		\fill[red!10] (2,1) rectangle ++(1,3);
		\draw[xscale=1,yscale=.5,step=1,gray,very thin] (0,0) grid (8,11);
		\draw[<->,thick] (0,5.5) node (yaxis) [above] {$t$} |- (8,0) node (xaxis) [right] {$r$};
		\foreach \i in {0,...,7}
		{
			\node at (.5+\i,-.3) {$\i$};
		}
		\foreach \j in {0,...,9}
		{
			\node at (-.25,.25+.5*\j) {$\j$};
		}
	\end{tikzpicture}
	\caption{Algorithm~\ref{alg:opt} when the initial point is not a preferred solution of a subproblem having less resource.}
	\label{fig:issue1}
\end{figure}
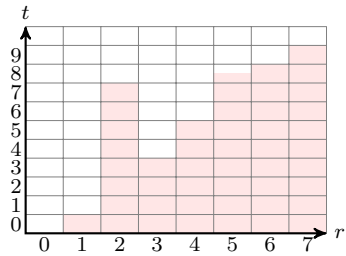

We first give two examples to illustrate that Algorithm~\ref{alg:opt} may not obtain a preferred solution from an arbitrary interior point.
Fig.~\ref{fig:issue1} illustrates the first example.
The algorithm only occupies new cells and it will not remove the already occupied ones no matter those occupied cells are the largest cells or not.
If the resource is exhausted while the occupied cells are not the largest ones, the output of the algorithm is not an optimal solution.

\begin{figure}
	\scriptsize
	\centering
	\begin{subfigure}{.25\textwidth}
		\centering
		\begin{tikzpicture}[scale=.5]
			\fill[red!10] (1,0) rectangle ++(1,.5);
			\fill[red!10] (2,0) rectangle ++(1,1);
			\fill[red!10] (4,0) rectangle ++(1,3);
			\fill[red!10] (6,0) rectangle ++(1,4.5);
			\fill[red!10] (7,0) rectangle ++(1,5);
			\fill[red!10] (5,0) rectangle ++(1,4.25);
			\fill[red!10] (3,0) rectangle ++(1,2);
			\fill[red!10] (2,1) rectangle ++(1,.25);
			\draw[xscale=1,yscale=.5,step=1,gray,very thin] (0,0) grid (8,11);
			\draw[<->,thick] (0,5.5) node (yaxis) [above] {$t$} |- (8,0) node (xaxis) [right] {$r$};
			\foreach \i in {0,...,7}
			{
				\node at (.5+\i,-.3) {$\i$};
			}
			\foreach \j in {0,...,9}
			{
				\node at (-.25,.25+.5*\j) {$\j$};
			}
		\end{tikzpicture}
		\caption{}
		\label{fig:issue2a}
	\end{subfigure}~~~~
	\begin{subfigure}{.25\textwidth}
		\centering
		\begin{tikzpicture}[scale=.5]
			\fill[red!10] (1,0) rectangle ++(1,.5);
			\fill[red!10] (2,0) rectangle ++(1,1);
			\fill[red!10] (4,0) rectangle ++(1,3);
			\fill[red!10] (6,0) rectangle ++(1,4.5);
			\fill[red!10] (7,0) rectangle ++(1,5);
			\fill[red!10] (5,0) rectangle ++(1,4.25);
			\fill[red!10] (3,0) rectangle ++(1,2);
			\fill[red!10] (2,1) rectangle ++(1,.25);
			\fill[red!20] (5,4.25) rectangle ++(1,.15);
			\draw[xscale=1,yscale=.5,step=1,gray,very thin] (0,0) grid (8,11);
			\draw[<->,thick] (0,5.5) node (yaxis) [above] {$t$} |- (8,0) node (xaxis) [right] {$r$};
			\foreach \i in {0,...,7}
			{
				\node at (.5+\i,-.3) {$\i$};
			}
			\foreach \j in {0,...,9}
			{
				\node at (-.25,.25+.5*\j) {$\j$};
			}
		\end{tikzpicture}
		\caption{}
		\label{fig:issue2b}
	\end{subfigure}
	\caption{Algorithm~\ref{alg:opt} when there are more than one non-integer $t_r$ at the initial point.}
	\label{fig:issue2}
\end{figure}
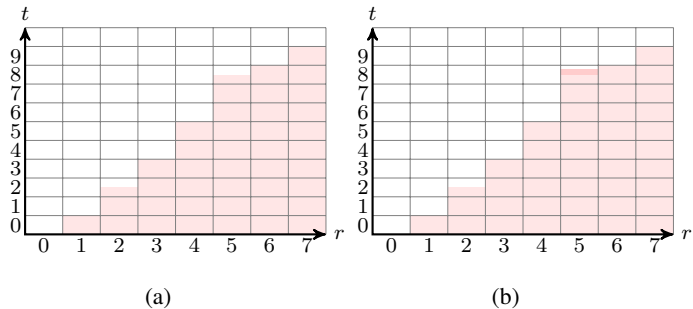

Fig.~\ref{fig:issue2a} illustrates another example where there are more than one non-integer $t_r$.
Suppose the resource is exhausted after filling a part of a partially filled cell as shown in Fig.~\ref{fig:issue2b}.
It is clear that the output is not a preferred solution.
However, if the two partially filled cells have the same value, the output may be an optimal solution.

Notice that Algorithm~\ref{alg:opt} is a greedy algorithm, so one way to understand the algorithm is to explore the optimal substructure of \eqref{eq:IP}.
We first define the subproblem of \eqref{eq:IP}.

\begin{definition}[Subproblem]
	An instance of \eqref{eq:IP} is a \emph{subproblem} of another instance of \eqref{eq:IP} if and only if the total resource of the former one is no more than that of the latter one.
\end{definition}

Recall in Fig.~\ref{fig:3dplot}, the height of the water surface is related to the allocated resource and the optimal objective under this amount of resource.
Therefore, the optimal solution of a subproblem has a higher water surface as there is less total resource.
As discussed in Section~\ref{sec:3d}, the algorithm lower the water surface in each iteration.
So, if we start from a preferred solution of a subproblem which can avoid the issue illustrated in Fig~\ref{fig:issue2}, e.g., $t_r = 0$ for all $r \in [M]$ and $0$ total resource, we can ensure that the algorithm outputs a preferred solution.

We now formally state the sufficient condition of obtaining a preferred solution by Algorithm~\ref{alg:opt}.

\begin{theorem} \label{thm:opt}
	Algorithm~\ref{alg:opt} outputs a feasible solution.
	If the input $\{t_r\}_{r = 0}^M$ is a preferred solution for a subproblem having less total resource where $t_r$ are all integers, then the output is also a preferred solution.
\end{theorem}

\begin{IEEEproof}
	See Appendix~\ref{sec:thm:opt}.
\end{IEEEproof}

In each iteration of Algorithm~\ref{alg:opt}, we need to find the $r$ which has the largest $\Delta_{r,\lfloor t_r \rfloor}$.
This can be handled by a max-heap, which can be initialized (heapified) in $\mathcal{O}(M)$ time.
Accessing such $r$ can be done in constant time.
After that, replacing the maximum value to $\Delta_{r,\lfloor t_r \rfloor+1}$ takes $\mathcal{O}(\log M)$ time (one deletion and one insertion).
Therefore, each iteration takes $\mathcal{O}(\log M)$ time.

The number of iterations in Algorithm~\ref{alg:opt} depends on the answer output by the algorithm.
Let $\{t_r^\ast\}_{r \in [M]}$ be the output of the algorithm and $\{\bar{t}_r\}_{r \in [M]}$ be the interior point input to the algorithm.
As the algorithm only increase the recoding number, there are totally $\lceil t_r^\ast \rceil - \lfloor \bar{t}_r \rfloor - \delta_{t_r^\ast, \bar{t}_r}$ iterations for each $r \in [M]$.
That is, the algorithm totally takes $\mathcal{O}(M+\sum_{r = 0}^M (\lceil t_r^\ast \rceil - \lfloor \bar{t}_r \rfloor - \delta_{t_r^\ast, \bar{t}_r}) \log M)$ time.
This complexity can be loosely upper bounded by $\mathcal{O}(M+\sum_{r = 0}^M \lceil t_r^\ast \rceil \log M)$.

If we can ensure that only the rank in the rank support $S$ would appear, we can set $\Delta_{r,t} = 0$ for all $r \not\in S$ and all $t \in \mathbb{N}$ so that we must have $t_r^\ast = 0$ for all $r \not\in S$.
Also, note that $t_0^\ast = 0$.
For simplicity, define $S^+ := S \setminus \{0\}$.
Then, we have 
\begin{equation*}
	\sum_{r = 0}^M (\lceil t_r^\ast \rceil - \lfloor \bar{t}_r \rfloor - \delta_{t_r^\ast, \bar{t}_r})
	\le \sum_{r \in S^+} \lceil t_r^\ast \rceil
	< \sum_{r \in S^+} t_r^\ast + |S^+| \le \frac{\sum_{r \in S^+} h_r t_r^\ast}{\min_{r \in S^+} h_r} + M
	= \frac{t_\text{avg}}{\min_{r \in S^+} h_r} + M.
\end{equation*}
This way, we can further simplify the complexity into $\mathcal{O}(M \log M + \frac{t_\text{avg}}{\min_{r \in S^+} h_r} \log M)$.

\subsection{From a Feasible Solution to a Preferred Solution}
\label{sec:primal2}

Suppose we have a feasible solution. %
This feasible solution may be the one obtained by Algorithm~\ref{alg:opt} from a ``bad'' interior point, or an optimal but not preferred solution output by other optimization problem solvers.
We now discuss the second algorithm, Algorithm~\ref{alg:tune}, which modifies the solution into a preferred one.

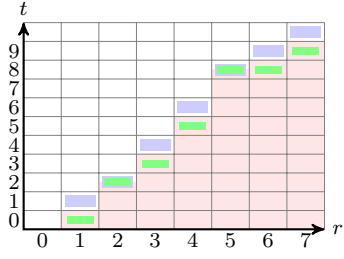
\begin{figure}
	\scriptsize
	\centering
	\begin{tikzpicture}[scale=.5]
		\fill[red!10] (1,0) rectangle ++(1,.5);
		\fill[red!10] (2,0) rectangle ++(1,1.25);
		\fill[red!10] (4,0) rectangle ++(1,3);
		\fill[red!10] (6,0) rectangle ++(1,4.5);
		\fill[red!10] (7,0) rectangle ++(1,5);
		\fill[red!10] (5,0) rectangle ++(1,4.25);
		\fill[red!10] (3,0) rectangle ++(1,2);
		\fill[blue!20] (1.1,.6) rectangle ++(.8,.3);
		\fill[blue!20] (2.1,1.1) rectangle ++(.8,.3);
		\fill[blue!20] (3.1,2.1) rectangle ++(.8,.3);
		\fill[blue!20] (4.1,3.1) rectangle ++(.8,.3);
		\fill[blue!20] (5.1,4.1) rectangle ++(.8,.3);
		\fill[blue!20] (6.1,4.6) rectangle ++(.8,.3);
		\fill[blue!20] (7.1,5.1) rectangle ++(.8,.3);
		\fill[green!50] (1.15,.15) rectangle ++(.7,.2);
		\fill[green!50] (2.15,1.15) rectangle ++(.7,.2);
		\fill[green!50] (3.15,1.65) rectangle ++(.7,.2);
		\fill[green!50] (4.15,2.65) rectangle ++(.7,.2);
		\fill[green!50] (5.15,4.15) rectangle ++(.7,.2);
		\fill[green!50] (6.15,4.15) rectangle ++(.7,.2);
		\fill[green!50] (7.15,4.65) rectangle ++(.7,.2);
		\draw[xscale=1,yscale=.5,step=1,gray,very thin] (0,0) grid (8,11);
		\draw[<->,thick] (0,5.5) node (yaxis) [above] {$t$} |- (8,0) node (xaxis) [right] {$r$};
		\foreach \i in {0,...,7}
		{
			\node at (.5+\i,-.3) {$\i$};
		}
		\foreach \j in {0,...,9}
		{
			\node at (-.25,.25+.5*\j) {$\j$};
		}
	\end{tikzpicture}
	\caption{Tuning from a feasible solution.}
	\label{fig:tune}
\end{figure}

The idea of the algorithm is that we release part of the smallest occupied cell and reallocate the resource to occupy the largest unoccupied cell if the objective value can be increased, which are the smallest of the green cells and the largest of the blue cells respectively in Fig.~\ref{fig:tune}.
Note that a partially filled cell can be considered as both green and blue.
By repeating this procedure, the algorithm eventually reaches an optimal solution, but such a solution is not necessary an ADS.
However, we know that excluding the partially filled cells, the remaining occupied cells are those largest cells.
Therefore, the last steps in the algorithm is to remove those partially filled cells so that we can obtain a preferred solution of a subproblem having equal or less resource.
This way, we can invoke Algorithm~\ref{alg:opt} to allocate the remaining resource and obtain a preferred solution.
Note that the algorithm has to access the value $\Delta_{r,-1}$ when $t_r = 0$.
To make sure that we do not output negative $t_r$, define $\Delta_{r,-1} := \infty$.

\begin{theorem} \label{thm:tune}
	The output of Algorithm~\ref{alg:tune} is a preferred solution.
\end{theorem}

\begin{IEEEproof}
	See Appendix~\ref{sec:thm:tune}.
\end{IEEEproof}

Similar to Algorithm~\ref{alg:opt}, each iteration of Algorithm~\ref{alg:tune} takes (twice) $\mathcal{O}(\log M)$ time by maintaining a max-heap and a min-heap.
The number of iterations depends on how far the optimal point is from the feasible point.
We consider the moment before the algorithm invoke Algorithm~\ref{alg:opt}.
Let $\{t_r^\ast\}_{r \in [M]}$ be the answer of the algorithm at this moment and $\{\bar{t}_r\}_{r \in [M]}$ be the input feasible solution.
Then, each $r \in [M]$ has $\max\{\lceil t_r^\ast \rceil - \lfloor \bar{t}_r \rfloor, \lceil \bar{t}_r \rceil - \lfloor t_r^\ast \rfloor\} - \delta_{t_r^\ast, \bar{t}_r}$ iterations.
Therefore, the algorithm takes $\mathcal{O}(M+\sum_{r = 0}^M (\max\{\lceil t_r^\ast \rceil - \lfloor \bar{t}_r \rfloor, \lceil \bar{t}_r \rceil - \lfloor t_r^\ast \rfloor\} - \delta_{t_r^\ast, \bar{t}_r}) \log M)$ time before invoking Algorithm~\ref{alg:opt}.
This complexity can be loosely upper bounded by $\mathcal{O}(M + \sum_{r = 0}^M \lceil \max\{t_r^\ast, \bar{t}_r\} \rceil \log M)$.

Again, if we can ensure that only the rank in the rank support $S$ would appear, we let $t_r^\ast = \bar{t}_r$ for all $r \not\in S^+$ as we can neglect these $r$ in the algorithm.
Then, we have
\begin{equation*}
	\sum_{r = 0}^M (\max\{\lceil t_r^\ast \rceil - \lfloor \bar{t}_r \rfloor, \lceil \bar{t}_r \rceil - \lfloor t_r^\ast \rfloor\} - \delta_{t_r^\ast, \bar{t}_r})
	\le \sum_{r \in S^+} \left\lceil \frac{t_\text{avg}}{h_r} \right\rceil
	\le \sum_{r \in S^+} \left( \frac{t_\text{avg}}{h_r} + 1 \right)
	\le \frac{Mt_\text{avg}}{\min_{r \in S^+} h_r} + M.
\end{equation*}
So, we can write the complexity before invoking Algorithm~\ref{alg:opt} as $\mathcal{O}(M \log M + \frac{Mt_\text{avg}}{\min_{r \in S^+} h_r} \log M)$.
Notice that the expression of the time complexity is the same after adding the one of Algorithm~\ref{alg:opt}.

\subsection{Tuning for Updated Parameters}
\label{sec:param}

We now consider that some parameters of \eqref{eq:IP} are updated.
Examples of parameters include the channel statistics (input rank distribution and the packet loss pattern on the outgoing channel), the batch size and the total resource.
The channel statistics may be learnt from the recently received batches or from the feedback of the network protocol.
The amendments of the batch size and the total resource are decided by other protocols or controlling algorithms which maintain the latency and buffer stability.

With both Algorithms~\ref{alg:opt} and \ref{alg:tune}, actually we can tune an arbitrary primal point $\{t_r\}_{r \in [M]}$ where $t_r \ge 0$ for all $r \in [M]$ to a preferred solution.
Such a primal point may be a preferred solution before the update of the parameters, which may become infeasible after the update.
Our guideline to tune an arbitrary primal point is as follows.
\begin{itemize}
	\item Case I: $\sum_{r = 0}^M h_r t_r = t_\text{avg}$. We can apply Algorithm~\ref{alg:tune} to obtain a preferred solution.
	\item Case II: $\sum_{r = 0}^M h_r t_r < t_\text{avg}$. We first apply Algorithm~\ref{alg:opt} to obtain a feasible solution, then apply Algorithm~\ref{alg:tune} to further tune the point into a preferred solution.
	\item Case III: $\sum_{r = 0}^M h_r t_r > t_\text{avg}$. This means that we need to release some occupied resource. After releasing, the point becomes a feasible (or an interior) point. Then, we can follow the above cases to obtain a preferred solution. One way to release the resource is to remove some smallest elements in $\biguplus_{r \in [M]} \Omega_r(t_r)$ so we only keep track of the larger elements, thus we can reduce the tuning steps induced by our algorithms.
\end{itemize}

If both the batch size $M$ and the outgoing channel statistic are not changed, we can further reduce the steps in the guideline when we tune an old preferred solution.
This is because the values of $\Delta_{r,t}$, i.e., the elements in $\biguplus_{r \in [M]} \Omega_r(\infty)$, remain the same and the old solution captures those largest elements in $\biguplus_{r \in [M]} \Omega_r(\infty)$.
In fact, this special scenario is the three cases we have discussed in detail in Section~\ref{sec:preferred}.

\subsection{Tuning from an Optimal Lagrange Multiplier}
\label{sec:dual}

Some solvers may apply a dual or a primal-dual approach to trickle the optimization problem.
That is, these solvers consider the dual problem \eqref{eq:D} and find an optimal Lagrange multiplier.
Binary search or subgradient search on the Lagrange multiplier are examples to find an optimal dual solution.
In this subsection, we discuss how to obtain a preferred solution from an optimal Lagrange multiplier.

An example of a standard dual-based solver is as follows.
There are two levels in the solver.
In the first level, we fix a $\lambda$ and then solve \eqref{eq:sup}, which is
	$\sup_{t_r \ge 0} ( h_r ( E_r(t_r) - \lambda t_r ) )$,
for each $r \in [M]$.
We can solve $t_r$ either by the solution shown in \eqref{eq:tr}, which is an interval for each $r$ and may need some extra search on the values of $\Delta_{r,t}$, or by using another way like subgradient search.
In the second level, given the $t_r$ solved in the first level, we try to adjust $\lambda$ to solve \eqref{eq:D}.
There are various ways such as binary search and subgradient search to update $\lambda$.
The two levels are applied alternatively.
This type of dual-based solver is globally asymptotically stable~\cite{protocol}.

We remark that in a subgradient search, there may be a sudden large jump when the new primal point lands on another line segment.
Therefore, we advise to use a diminishing step size for a faster convergence.

Suppose we use \eqref{eq:tr} to solve $t_r$.
If $\sum_{r = 1}^M h_r t_r - t_\text{avg}$ does not have an opposite sign for all $t_r$ in the interval shown in \eqref{eq:tr}, then it is safe to choose any $t_r$ in that interval to be the primal values used for the iteration in the second level.
Otherwise, it means that the optimal $t_r$ is laid on the interval shown in \eqref{eq:tr}, i.e., an optimal $\lambda$ is found.

Regardless of how we find an optimal $\lambda$, we need an extra step to ensure that the primal solution is a preferred solution.
As \eqref{eq:tr} is independent of $(h_0, \ldots, h_M)$, we can set $t_r = \min t_r(\lambda)$ for all $r$ to be a preferred solution of a subproblem having less resource.
It is a preferred solution as $\biguplus_{r \in [M]} \Omega_r(\min t_r(\lambda))$ is a collection of all elements in $\biguplus_{r \in [M]} \Omega_r(\infty)$ which are smaller than $\lambda$; and $\min t_r(\lambda)$ are integers for all $r \in [M]$.
Then, according to Theorem~\ref{thm:opt}, we can apply Algorithm~\ref{alg:opt} to obtain a preferred solution.

\section{Concluding Remarks} \label{sec:conclude}

We proposed a general recoding number optimization problem for batched network coding with adaptive recoding, studied the properties of the optimization problem, and provided algorithms to obtain a preferred solution which can tolerate rank distribution errors.
Our approach works for a very general setting: arbitrary finite field for recoding and packet loss pattern subject to the concavity of the expected rank functions.
We also viewed the problem from different angles, showed an almost deterministic property on the optimal recoding numbers, and proved an intuition that the recoding number of a batch of a larger rank should be no smaller than that of a batch of a lower rank.
These investigations can simplify the analysis and design of other applications based on adaptive recoding.

To further generalize our framework, it is interesting to know whether the concavity of the expected rank functions can be preserved when we apply %
systematic recoding \cite{yang14a,bats_book}. %
On the other hand, our example of non-stationary loss pattern that satisfies the concavity requirement suggests that there may be a more general class of loss patterns guaranteeing the concavity of the expected rank functions which is worth to study.
It is also important to investigate the gap from optimality of our algorithms when the expected rank functions are not concave.

Our approach can be applied to a general batched network code.
But as we use the expected rank as the recoding objective, our approach in general assumes that the outer code can achieve the expected rank of batch transfer matrices.
Notice that not all batched network codes studied in the literature can achieve the expected rank.
For example, when the batches are generated using disjoint subsets of input packets, the decoding requires that the rank of each batch is larger than a certain value.
For this case, the objective of our optimization should be changed.
One direction is to introduce a penalty function to the objective which emulates the throughput of the code.
Another way is to modify the physical meaning of the expected rank functions to match with the throughput of the code, but this may breach the concavity.

As baseline recoding maintains certain stochastic dominance relations \cite{variable}, we suspect that adaptive recoding would also do so.
This may be the key to investigate the relation between the distributed model at each node and the centralized model of the whole network.
We may need to reformulate the problem as a stochastic optimization problem instead of maximizing the average expected rank at the next node.
This is another future direction which aims to understand the stochastic nature of adaptive recoding.

\newpage

\appendices

\section{Proof of Lemma~\ref{lem:ert_matrix}} \label{sec:lem:ert_matrix}

Let $\{\mathbf{v}_1, \ldots, \mathbf{v}_r, \ldots, \mathbf{v}_s\}$ be the received packets in a batch.
Without loss of generality, let $\{\mathbf{v}_1, \ldots, \mathbf{v}_r\}$ be a basis of the vector space spanned by all the received packets in the batch.
Then, we have
\begin{equation*}
	\mathbf{v}_j = \sum_{\substack{i = 1\\c_{i,j} \neq 0}}^r c_{i,j} \mathbf{v}_i
\end{equation*}
for all $j = r+1, \ldots, s$ and some constants $c_{i,j} \in \mathbb{F}_q$.
A recoded packet generated by RLNC is a vector expressed by $\sum_{k = 1}^s \beta_k \mathbf{v}_k = \sum_{k = 1}^r \beta'_k \mathbf{v}_k$,
where $\beta_1, \ldots, \beta_s$ are chosen from $\mathbb{F}_q$ uniformly, and
\begin{equation*}
	\beta'_k = \beta_k + \sum_{\substack{\ell = r+1\\c_{k,\ell} \neq 0}}^s \beta_\ell c_{k,\ell}.
\end{equation*}

Let $\beta, \gamma$ be random variables which are uniformly distributed over all entries in $\mathbb{F}_q$.
For all $a, c \in \mathbb{F}_q$ where $c \neq 0$, we have $\Pr(\beta c = a) = \Pr(\beta = a c^{-1}) = 1/q$.
On the other hand, every row in the addition table of $\mathbb{F}_q$ is a permutation of all the entries in $\mathbb{F}_q$.
So, we have $\Pr(\beta+\gamma = a) = q (1/q)^2 = 1/q$.
The above arguments show that $\beta'_k$ is distributed uniformly over $\mathbb{F}_q$.
That is, the RLNC of all received packets is equivalent in distribution to the RLNC of all linearly independent received packets.

Let $\mathbf{B}$ be the matrix formed by juxtaposing $\{\mathbf{v}_1, \ldots, \mathbf{v}_r\}$.
The recoded packets are the columns in the matrix $\mathbf{B}\mathbf{R}_{r,t}$.
Let $\mathbf{T}$ be the matrix formed by juxtaposing the received recoded packets at the next hop.
Note that $\rk(\mathbf{T}) = \rk(\mathbf{B}\mathbf{R}_{r,t}\mathbf{Z}_t)$, where the latter substitutes the lost packets into zero column vectors.
So, we can express the expected rank function by
\begin{equation*}
	E_r(t) = \mathbb{E}[\rk(\mathbf{B}\mathbf{R}_{r,t}\mathbf{Z}_t)] = \mathbb{E}[\rk(\mathbf{R}_{r,t}\mathbf{Z}_t)],
\end{equation*}
where the last equality holds as $\mathbf{B}$ has full column rank.

\section{Proof of Lemma~\ref{lem:concaveR}} \label{sec:lem:concaveR}

Without loss of generality, we can write $\mathbf{R}_{r,t+1}=(\mathbf{R}_{r,t}\mid \mathbf{x})$ and $\mathbf{R}_{r,t+2}=(\mathbf{R}_{r,t}\mid \mathbf{x}\mid \mathbf{y})$, where $\mathbf{x}$ and $\mathbf{y}$ are independent totally random column vectors.

When $r = 0$, both $\mathbf{R}_{r,t}$ and $\mathbf{R}_{r,t+1}$ are empty matrices.
By definition, their ranks are both $0$.
So, we have $\mathbb{E}[\rk(\mathbf{R}_{0,t})] = 0$ for all $t$.

Now we consider $r > 0$.
Denoted by $\Span(\mathbf{M})$ the vector space spanned by the columns of a matrix $\mathbf{M}$.
We have
\begin{IEEEeqnarray*}{rCl}
	\Delta \mathbb{E}[\rk(\mathbf{R}_{r,t})]&=&\mathbb{E}[\rk(\mathbf{R}_{r,t+1})-\rk(\mathbf{R}_{r,t})]\\
	&=& \Pr(\rk(\mathbf{R}_{r,t+1})-\rk(\mathbf{R}_{r,t})=1) \yesnumber \label{eq:rankdiff} \\
	&=& \Pr(\mathbf{x} \not \in \Span(\mathbf{R}_{r,t})),
\end{IEEEeqnarray*}
where \eqref{eq:rankdiff} holds since $\rk(\mathbf{R}_{r,t+1}) - \rk(\mathbf{R}_{r,t})$ is a binary random variable.
Similarly, we have
\begin{equation*}
	\Delta \mathbb{E}[\rk(\mathbf{R}_{r,t+1})] = \Pr(\mathbf{y} \not \in \Span(\mathbf{R}_{r,t}\mid \mathbf{x})).
\end{equation*}

On the other hand,
\begin{IEEEeqnarray}{rCl}
	\Delta \mathbb{E}[\rk(\mathbf{R}_{r,t})] & = & \Pr(\mathbf{x} \not \in \Span(\mathbf{R}_{r,t})) \nonumber \\
	& = & \Pr(\mathbf{y} \not \in \Span(\mathbf{R}_{r,t})) \label{eq:lem2span1} \\
	& \ge & \Pr(\mathbf{y} \not \in \Span(\mathbf{R}_{r,t}\mid \mathbf{x})) \label{eq:lem2span2} \\
	& = & \Delta \mathbb{E}[\rk(\mathbf{R}_{r,t+1})], \nonumber
\end{IEEEeqnarray}
where
\begin{itemize}
	\item \eqref{eq:lem2span1} holds since $\mathbf{x}$ and $\mathbf{y}$ are equiprobable and independent of $\mathbf{R}_{r,t}$; and
	\item \eqref{eq:lem2span2} holds since $\Span(\mathbf{R}_{r,t}) \subseteq \Span(\mathbf{R}_{r,t} \mid \mathbf{x})$.
\end{itemize}
This also shows that $\mathbb{E}[\rk(\mathbf{R}_{r,t})]$ is monotonic increasing and concave with respect to $t$.

When $q$ is finite, we know that
\ifnum\paperversion=1
\begin{multline*}
	\Pr(\Span(\mathbf{R}_{r,t}) \neq \Span(\mathbf{R}_{r,t} \mid \mathbf{x}))\\
	= \Pr(\mathbf{x} \not \in \Span(\mathbf{R}_{r,t})) > 0.
\end{multline*}
\else
\begin{equation*}
	\Pr(\Span(\mathbf{R}_{r,t}) \neq \Span(\mathbf{R}_{r,t} \mid \mathbf{x})) = \Pr(\mathbf{x} \not \in \Span(\mathbf{R}_{r,t})) > 0.
\end{equation*}
\fi
As $\mathbf{y}$ is also independent of $\mathbf{R}_{r,t}$ and $\mathbf{x}$, it is straightforward to see that $\Pr(y \not \in \Span(\mathbf{R}_{r,t} \mid \mathbf{x})) > \Pr(y \not \in \Span(\mathbf{R}_{r,t}))$, i.e., the inequality in \eqref{eq:lem2span2} is strict.

Now we consider $q \to \infty$.
We know that $\mathbf{y} \not \in \Span(\mathbf{R}_{r,t})$ with probability tends to $1$ when $0 \le t < r$; and almost never when $t \ge r > 0$.
That is, the inequality in \eqref{eq:lem2span2} is strict when $t+1 = r > 0$.
As we are considering the points $\{t, t+1, t+2\}$, it means that $\mathbb{E}[\rk(\mathbf{R}_{r,t})]$ is strictly concave at $r$.

The proof of (\ref{lem:concaveR:a}) is done.
For (\ref{lem:concaveR:b}), let $\tilde{\mathbf{x}} = (\mathbf{x}^T \mid z)^T$, where $z$ is a single element uniformly drawn from the underlying field.

If $\tilde{\mathbf{x}} \in \Span(\mathbf{R}_{r+1,t})$, then we can write $\tilde{\mathbf{x}}$ as a linear combination of the columns of $\mathbf{R}_{r+1,t}$.
By removing the last row, we can show that $\mathbf{x}$ is a linear combination of the columns of $\mathbf{R}_{r,t}$.
That is, $\mathbf{x} \in \Span(\mathbf{R}_{r,t})$.
So, we have
\begin{equation*}
	\Pr(\tilde{\mathbf{x}} \in \Span(\mathbf{R}_{r+1,t})) \le \Pr(\mathbf{x} \in \Span(\mathbf{R}_{r,t})),
\end{equation*}
where the inequality is strict if $q$ is finite.

On the other hand, recall that
\begin{equation*}
	\Delta \mathbb{E}[\rk(\mathbf{R}_{r,t})] = \begin{cases}
		\Pr(\mathbf{x} \not \in \Span(\mathbf{R}_{r,t})) & \text{if } r > 0,\\
		0 & \text{otherwise},
	\end{cases}
\end{equation*}
and
\begin{equation*}
	\Delta \mathbb{E}[\rk(\mathbf{R}_{r+1,t})] = \Pr(\tilde{\mathbf{x}} \not \in \Span(\mathbf{R}_{r+1,t})).
\end{equation*}
So, we have $\Delta \mathbb{E}[\rk(\mathbf{R}_{r+1,t})] \ge \Delta \mathbb{E}[\rk(\mathbf{R}_{r,t})]$.
The inequality is strict if $q$ is finite.

For $q \to \infty$, we have $\rk(\mathbf{R}_{r,t}) = \min\{r,t\}$ with probability tends to $1$.
So, we have
\begin{equation*}
	\Delta\mathbb{E}[\rk(\mathbf{R}_{r,t})] = \begin{cases}
		1 & \text{if } t < r,\\
		0 & \text{otherwise}.
	\end{cases}
\end{equation*}
It is easy to see that $\Delta \mathbb{E}[\rk(\mathbf{R}_{r+1,t})] > \Delta \mathbb{E}[\rk(\mathbf{R}_{r,t})]$ when $r = t$.

The proof of (\ref{lem:concaveR:b}) is done.
Thus, we have finished proving Lemma~\ref{lem:concaveR}.

\section{Proof of Theorem~\ref{thm:concave}} \label{sec:thm:concave}

We know that $\{Z_t\}$ is a stationary stochastic process \cite{protocol}.
Recall that the columns in $\mathbf{R}_{r,t}$ have the same distribution.
Under the condition that the number of $1$ in $\mathbf{Z}_t$ equals $i$, $\rk(\mathbf{R}_{r,t}\mathbf{Z}_t)$ has the same distribution as $\rk(\mathbf{R}_{r,i})$.
\ifnum\paperversion=1
The calculation to show the concavity is shown at the top of the next page, where
\else
Consider
	\begin{IEEEeqnarray*}{Cl}
		& E_r(t+2) - E_r(t+1) = \mathbb{E}[\rk(\mathbf{R}_{r,t+2}\mathbf{Z}_{t+2}) - \rk(\mathbf{R}_{r,t+1}\mathbf{Z}_{t+1})] \yesnumber \label{eq:concave2} \\
		= & \sum_{i = 0}^{t+1} \left\{ \Pr(Z_{t+2} = 1, \rk(\mathbf{Z}_{t+1}) = i) \mathbb{E}[\rk(\mathbf{R}_{r,i+1})-\rk(\mathbf{R}_{r,i})] \right.\\
		& \qquad \left. + \Pr(Z_{t+2} = 0, \rk(\mathbf{Z}_{t+1}) = i) \mathbb{E}[\rk(\mathbf{R}_{r,i})-\rk(\mathbf{R}_{r,i})] \right\}\\
		= & \sum_{i = 1}^{t+1} \Pr( Z_1 = 1, Z_{t+2} = 1, \rk(\mathbf{Y}_{t+1}) = i-1) \Delta \mathbb{E}[\rk(\mathbf{R}_{r,i})]\\
		& \qquad + \sum_{i = 0}^t \Pr( Z_1 = 0, Z_{t+2} = 1, \rk(\mathbf{Y}_{t+1}) = i) \Delta \mathbb{E}[\rk(\mathbf{R}_{r,i})]\\
		= & \sum_{i = 0}^t \left\{ \Pr( Z_1 = 1, Z_{t+2} = 1, \rk(\mathbf{Y}_{t+1}) = i) \Delta \mathbb{E}[\rk(\mathbf{R}_{r,i+1})] \right.\\
		& \qquad \left. + \Pr( Z_1 = 0, Z_{t+2} = 1, \rk(\mathbf{Y}_{t+1}) = i) \Delta \mathbb{E}[\rk(\mathbf{R}_{r,i})] \right\} \yesnumber \label{eq:concave3} \\
		\le & \sum_{i = 0}^t \Pr( Z_{t+2} = 1, \rk(\mathbf{Y}_{t+1}) = i) \Delta \mathbb{E}[\rk(\mathbf{R}_{r,i})] \yesnumber \label{eq:concave3a} \\
		= & \sum_{i = 0}^t \Pr( Z_{t+1} = 1, \rk(\mathbf{Z}_t) = i) \Delta \mathbb{E}[\rk(\mathbf{R}_{r,i})] \yesnumber \label{eq:concave4}\\
		= & \sum_{i = 0}^t \left\{ \Pr(Z_{t+1} = 1, \rk(\mathbf{Z}_t) = i) \mathbb{E}[\rk(\mathbf{R}_{r,i+1})-\rk(\mathbf{R}_{r,i})] \right.\\
		& \qquad \left. + \Pr(Z_{t+1} = 0, \rk(\mathbf{Z}_t) = i) \mathbb{E}[\rk(\mathbf{R}_{r,i})-\rk(\mathbf{R}_{r,i})] \right\}\\
		= & \mathbb{E}[\rk(\mathbf{R}_{r,t+1}\mathbf{Z}_{t+1}) - \rk(\mathbf{R}_{r,t}\mathbf{Z}_t)] = E_r(t+1) - E_r(t), \yesnumber \label{eq:concave5}
	\end{IEEEeqnarray*}
	where
\fi
\begin{itemize}
	\item \eqref{eq:concave2} and \eqref{eq:concave5} are according to Lemma~\ref{lem:ert_matrix};
\ifnum\paperversion=1
	\item \eqref{eq:concave4} holds by applying Lemma~\ref{lem:concaveR}(\ref{lem:concaveR:a}) and the assumption that $\{Z_t\}$ is a stationary stochastic process.
\else
	\item \eqref{eq:concave3a} holds by applying Lemma~\ref{lem:concaveR}(\ref{lem:concaveR:a});
	\item \eqref{eq:concave4} by the assumption that $\{Z_t\}$ is a stationary stochastic process.
\fi
\end{itemize}

\ifnum\paperversion=1
\begin{figure*}
	\begin{IEEEeqnarray*}{Cl}
		& E_r(t+2) - E_r(t+1) = \mathbb{E}[\rk(\mathbf{R}_{r,t+2}\mathbf{Z}_{t+2}) - \rk(\mathbf{R}_{r,t+1}\mathbf{Z}_{t+1})] \yesnumber \label{eq:concave2} \\
		= & \sum_{i = 0}^{t+1} \left\{ \Pr(Z_{t+2} = 1, \rk(\mathbf{Z}_{t+1}) = i) \mathbb{E}[\rk(\mathbf{R}_{r,i+1})-\rk(\mathbf{R}_{r,i})] + \Pr(Z_{t+2} = 0, \rk(\mathbf{Z}_{t+1}) = i) \mathbb{E}[\rk(\mathbf{R}_{r,i})-\rk(\mathbf{R}_{r,i})] \right\}\\
		= & \sum_{i = 1}^{t+1} \Pr( Z_1 = 1, Z_{t+2} = 1, \rk(\mathbf{Y}_{t+1}) = i-1) \Delta \mathbb{E}[\rk(\mathbf{R}_{r,i})] + \sum_{i = 0}^t \Pr( Z_1 = 0, Z_{t+2} = 1, \rk(\mathbf{Y}_{t+1}) = i) \Delta \mathbb{E}[\rk(\mathbf{R}_{r,i})]\\
		= & \sum_{i = 0}^t \left\{ \Pr( Z_1 = 1, Z_{t+2} = 1, \rk(\mathbf{Y}_{t+1}) = i) \Delta \mathbb{E}[\rk(\mathbf{R}_{r,i+1})] + \Pr( Z_1 = 0, Z_{t+2} = 1, \rk(\mathbf{Y}_{t+1}) = i) \Delta \mathbb{E}[\rk(\mathbf{R}_{r,i})] \right\} \yesnumber \label{eq:concave3} \\
		\le & \sum_{i = 0}^t \Pr( Z_{t+2} = 1, \rk(\mathbf{Y}_{t+1}) = i) \Delta \mathbb{E}[\rk(\mathbf{R}_{r,i})] = \sum_{i = 0}^t \Pr( Z_{t+1} = 1, \rk(\mathbf{Z}_t) = i) \Delta \mathbb{E}[\rk(\mathbf{R}_{r,i})] \yesnumber \label{eq:concave4}\\
		= & \sum_{i = 0}^t \left\{ \Pr(Z_{t+1} = 1, \rk(\mathbf{Z}_t) = i) \mathbb{E}[\rk(\mathbf{R}_{r,i+1})-\rk(\mathbf{R}_{r,i})] + \Pr(Z_{t+1} = 0, \rk(\mathbf{Z}_t) = i) \mathbb{E}[\rk(\mathbf{R}_{r,i})-\rk(\mathbf{R}_{r,i})] \right\}\\
		= & \mathbb{E}[\rk(\mathbf{R}_{r,t+1}\mathbf{Z}_{t+1}) - \rk(\mathbf{R}_{r,t}\mathbf{Z}_t)] = E_r(t+1) - E_r(t). \yesnumber \label{eq:concave5}
	\end{IEEEeqnarray*}
	\hrule
\end{figure*}
\fi

We can see from \eqref{eq:concave5} that $E_r(t)$ is monotonic increasing as the rank of $\mathbf{R}_{r,t+1}\mathbf{Z}_{t+1}$ is no less than the rank of $\mathbf{R}_{r,t}\mathbf{Z}_t$ due to coupling.

Now, we consider $t+1 = c$.
Note that $\Pr(Z_1 = 1, Z_{c+1} = 1) \neq 0$ implies that $\Pr(Z_1 = 1, Z_{t+2} = 1, \rk(\mathbf{Y}_{t+1}) = i) \neq 0$ for some $0 \le i \le t$.
When $q$ is finite, we have $\Delta\mathbb{E}[\rk(\mathbf{R}_{r,i+1})] < \Delta\mathbb{E}[\rk(\mathbf{R}_{r,i})]$ for all $r > 0$ by Lemma~\ref{lem:concaveR}(\ref{lem:concaveR:a}).
In this case, the inequality in \eqref{eq:concave4} is strict if $r > 0$, i.e., $E_r(t)$ is strictly concave at $c$ under this condition.

When $q \to \infty$, Lemma~\ref{lem:concaveR}(\ref{lem:concaveR:a}) tells us that $\Delta\mathbb{E}[\rk(\mathbf{R}_{r,i+1})] = \Delta\mathbb{E}[\rk(\mathbf{R}_{r,i})]$ for all $i+1 \neq r > 0$.
So, we only need to consider $i+1 = r > 0$.
That is, we need $\Pr(Z_1 = 1, Z_{c+1} = 1, \rk(\mathbf{Y}_c) = r-1) \neq 0$ to achieve a strict inequality.

\section{Proof of Theorem~\ref{thm:deltar}} \label{sec:thm:deltar}

Under the condition that the number of $1$ in $\mathbf{Z}_t$ equals $i$, $\rk(\mathbf{R}_{r,t}\mathbf{Z}_t)$ has the same distribution as $\rk(\mathbf{R}_{r,i})$.
We consider the following:
\begin{IEEEeqnarray*}{Cl}
	& \Delta_{r+1,t} = E_{r+1}(t+1) - E_{r+1}(t)\\
	= & \mathbb{E}[\rk(\mathbf{R}_{r+1,t+1}\mathbf{Z}_{t+1}) - \rk(\mathbf{R}_{r+1,t}\mathbf{Z}_t)] \yesnumber \label{eq:deltart1} \\
	= & \sum_{i = 0}^t \Pr(Z_{t+1} = 1, \rk(\mathbf{Z}_t) = i) \Delta\mathbb{E}[\rk(\mathbf{R}_{r+1,i})]\\
	\ge & \sum_{i = 0}^t \Pr(Z_{t+1} = 1, \rk(\mathbf{Z}_t) = i) \Delta\mathbb{E}[\rk(\mathbf{R}_{r,i})] \yesnumber \label{eq:deltart2} \\
	= & \mathbb{E}[\rk(\mathbf{R}_{r,t+1}\mathbf{Z}_{t+1}) - \rk(\mathbf{R}_{r,t}\mathbf{Z}_t)]\\
	= & E_r(t+1) - E_r(t) = \Delta_{r,t}, \yesnumber \label{eq:deltart3}
\end{IEEEeqnarray*}
	where
	\begin{itemize}
		\item \eqref{eq:deltart1} and \eqref{eq:deltart3} are by Lemma~\ref{lem:ert_matrix};
		\item \eqref{eq:deltart2} is by Lemma~\ref{lem:concaveR}(\ref{lem:concaveR:b}).
	\end{itemize}

	For finite $q$, Lemma~\ref{lem:concaveR}(\ref{lem:concaveR:b}) states that $\Delta\mathbb{E}[\rk(\mathbf{R}_{r,i})]$ is strictly increasing with respect to $r$.
	Note that $\Pr(Z_{t+1} = 1) \neq 0$ implies that $\Pr(Z_{t+1}, \rk(\mathbf{Z}_t) = i) \neq 0$ for some $0 \le i \le t$, which shows that the inequality in \eqref{eq:deltart2} is strict.

	When $q \to \infty$, Lemma~\ref{lem:concaveR}(\ref{lem:concaveR:b}) states that $\Delta\mathbb{E}[\rk(\mathbf{R}_{r,i})]$ is strictly increasing with respect to $r$ only at $r = i$.
	To make a strict inequality in \eqref{eq:deltart2}, we require $\Pr(Z_{t+1} = 1, \rk(\mathbf{Z}_t) = r) \neq 0$.

\section{Proof of Theorem~\ref{thm:diag}} \label{sec:thm:diag}

We need to evaluate $\Delta_{r+1,t+1} = E_{r+1}(t+2) - E_{r+1}(t+1)$.
We can follow the same steps from \eqref{eq:concave2} to \eqref{eq:concave3} except substituting $r$ into $r+1$ to obtain
\begin{align*}
	\sum_{i = 0}^t \{ \Pr( Z_1 = 1, Z_{t+2} = 1, \rk(\mathbf{Y}_{t+1}) = i) \Delta \mathbb{E}[\rk(\mathbf{R}_{r+1,i+1})]\\
	+\; \Pr( Z_1 = 0, Z_{t+2} = 1, \rk(\mathbf{Y}_{t+1}) = i) \Delta \mathbb{E}[\rk(\mathbf{R}_{r+1,i})] \}. \stepcounter{equation} \tag{\theequation} \label{eq:diag1}
\end{align*}

By \eqref{eq:diag0} and Lemma~\ref{lem:concaveR}(\ref{lem:concaveR:b}), we know that \eqref{eq:diag1} is larger than or equal to
\begin{equation} \label{eq:diag2}
	\sum_{i = 0}^t \Pr( Z_{t+2} = 1, \rk(\mathbf{Y}_{t+1}) = i) \Delta \mathbb{E}[\rk(\mathbf{R}_{r,i})].
\end{equation}

When $q$ is finite, by applying Lemma~\ref{lem:concaveR}(\ref{lem:concaveR:b}), we can see that the inequality in \eqref{eq:diag2} is strict when $\Pr(Z_1 = 0, Z_{t+2} = 1, \rk(\mathbf{Y}_{t+1}) = i) \neq 0$ for some $0 \le i \le t$.
This condition can be covered by $\Pr(Z_1 = 0, Z_{t+2} = 1) \neq 0$.

When $q \to \infty$, Lemma~\ref{lem:concaveR}(\ref{lem:concaveR:b}) tells us that we only need to consider $r = i$.
We require $\Pr(Z_1 = 0, Z_{t+2} = 1, \rk(\mathbf{Y}_{t+1}) = r) \neq 0$ to make a strict inequality in \eqref{eq:diag2}.

Note that \eqref{eq:diag2} is exactly the same as \eqref{eq:concave4} because $\{Z_t\}$ is a stationary stochastic process.
So, we can follow the steps from \eqref{eq:concave4} to \eqref{eq:concave5} and finish the proof.

\section{Proof of Theorem~\ref{thm:integer}} \label{sec:thm:integer}

Recall that $S$ denotes the rank support (see Definition~\ref{def:ranksupport}).

	For those $r \in [M] \setminus S$, we can set $t_r = 0$ as $h_r E_r(\cdot) = 0$.
	Similarly, we can set $t_0 = 0$ as $E_0(\cdot) = 0$.

	Now we consider $r \in S$.
	If $|S| = 1$, the proof is done.
	Suppose there are distinct $r, r' \in S$ such that $t_r$ and $t_{r'}$ are non-integers.
	Our goal is to show that eliminating the fractional part of $t_r$ or $t_{r'}$ gives a non-decreased objective value.

	The sum of the expected ranks (at the next hop) for the batches of rank $r$ and $r'$ is
	\begin{equation} \label{eq:2e}
		h_r E_r(t_r) + h_{r'} E_{r'}(t_{r'}).
	\end{equation}
	Without loss of generality, assume $\Delta_{r, t_r} \ge \Delta_{r', t_{r'}}$.
	Let $\epsilon_r = t_r - \lfloor t_r \rfloor$ and $\epsilon_{r'} = t_{r'} - \lfloor t_{r'} \rfloor$.

	Case I: $h_{r'}\epsilon_{r'} \ge h_r(1-\epsilon_r)$.
	We move a portion of $\epsilon_{r'}$ to $t_r$ so that $t_r$ becomes $\lfloor t_r \rfloor + 1$, which is an integer.
	In order to keep the total resource used by $r$ and $r'$ constant, $t_{r'}$ is decreased by $\frac{h_r(1-\epsilon_r)}{h_{r'}}$.
	Then the sum of the expected ranks becomes
	\begin{equation} \label{eq:2ecase1}
		h_r E_r(\lfloor t_r \rfloor + 1) + h_{r'} E_{r'}\left(t_{r'}-\frac{h_r(1-\epsilon_r)}{h_{r'}}\right),
	\end{equation}
	and \eqref{eq:2ecase1} minus \eqref{eq:2e} gives $h_r(1-\epsilon_r)(\Delta_{r,t_r}-\Delta_{r',t_{r'}})$, which is non-negative.

	Case II: $h_{r'}\epsilon_{r'} < h_r(1-\epsilon_r)$.
	We move all $\epsilon_{r'}$ to $t_r$ so that $t_{r'}$ becomes an integer.
	To balance the total resource used by $r$ and $r'$, $t_r$ is increased by $\frac{h_{r'}\epsilon_{r'}}{h_r}$.
	Note that $t_r+\frac{h_{r'}\epsilon_{r'}}{h_r} < \lfloor t_r \rfloor + 1$.
	Then the sum of expected ranks becomes
	\begin{equation} \label{eq:2ecase2}
		h_r E_r\left(t_r+\frac{h_{r'}\epsilon_{r'}}{h_r}\right) + h_{r'} E_{r'}(\lfloor t_{r'} \rfloor),
	\end{equation}
	and \eqref{eq:2ecase2} minus \eqref{eq:2e} gives $h_{r'} \epsilon_{r'} (\Delta_{r,t_r}-\Delta_{r',t_{r'}})$, which is non-negative.

	Combine the two cases, we can reduce the number of non-integer $t_r$ without decreasing the objective value.
	By applying the procedure repeatedly, the proof is done.

\section{Proof of Theorem~\ref{thm:IPomega}} \label{sec:thm:IPomega}

	Suppose $\biguplus_{r \in S} \Omega_r(t_r)$ is not a collection of the largest $\sum_{r \in S} \lceil t_r \rceil$ elements in $\biguplus_{r \in S} \Omega_r(\infty)$.
	Then, there exists distinct $m, n \in S$ such that
	\begin{equation} \label{eq:m>n}
		\Omega_m(t_m) \not \ni \Delta_{m, \lceil t_m \rceil} > \Delta_{n, \lceil t_n - 1 \rceil} \in \Omega_n(t_n).
	\end{equation}

	The resource we need to increase $t_m$ into the next larger integer $\lfloor t_m + 1 \rfloor$ is $h_m (1 - (t_m - \lfloor t_m \rfloor))$.
	On the other hand, the resource we release by decreasing $t_n$ to the next smaller integer $\lceil t_n - 1 \rceil$ is $h_n (t_n - \lfloor t_n \rfloor + \delta_{t_n,\lfloor t_n \rfloor})$.
	Define
	\begin{equation*}
		s = \min\{h_m (1 - (t_m - \lfloor t_m \rfloor)), h_n (t_n - \lfloor t_n \rfloor + \delta_{t_n,\lfloor t_n \rfloor})\} > 0.
	\end{equation*}

	We reallocate $s$ amount of resource between $t_m$ and $t_n$.
	Let
	\begin{equation} \label{eq:t'_r}
		t'_r = \begin{cases}
			t_m + s/h_m \le \lfloor t_m + 1 \rfloor & \text{if } r = m,\\
			t_n - s/h_n \ge \lceil t_n - 1 \rceil & \text{if } r = n,\\
			t_r & \text{otherwise}.
		\end{cases}
	\end{equation}

	As we only reallocate the resource, $\{t'_r\}_{r = 0}^M$ is obviously a feasible solution of \eqref{eq:IP}.
	Now we consider the change in the objective value:
	\ifnum\paperversion=1
	\begin{IEEEeqnarray*}{Cl}
		& \sum_{r = 0}^M h_r E_r(t'_r) - \sum_{r = 0}^M h_r E_r(t_r)\\
		= & s(\Delta_{m, \lfloor t_m \rfloor} - \Delta_{n, \lceil t_n - 1 \rceil})\\
		\ge & s(\Delta_{m, \lceil t_m \rceil} - \Delta_{n, \lceil t_n - 1 \rceil}) \yesnumber \label{eq:floor>ceil} \\
		> & 0 \yesnumber \label{eq:floor>ceil>0},
	\end{IEEEeqnarray*}
	\else
	\begin{IEEEeqnarray*}{rCl}
		\sum_{r = 0}^M h_r E_r(t'_r) - \sum_{r = 0}^M h_r E_r(t_r)
		& = & s(\Delta_{m, \lfloor t_m \rfloor} - \Delta_{n, \lceil t_n - 1 \rceil})\\
		& \ge & s(\Delta_{m, \lceil t_m \rceil} - \Delta_{n, \lceil t_n - 1 \rceil}) \yesnumber \label{eq:floor>ceil} \\
		& > & 0 \yesnumber \label{eq:floor>ceil>0},
	\end{IEEEeqnarray*}
	\fi
	where \eqref{eq:floor>ceil} holds as $\Delta_{m, \lfloor t_m \rfloor} \ge \Delta_{m, \lceil t_m \rceil}$, and \eqref{eq:floor>ceil>0} is followed by \eqref{eq:m>n}.

	This means that $\{t_r\}_{r = 0}^M$ is not an optimal solution.
	The proof is done by contraposition.

\section{Proof of Theorem~\ref{thm:omegaIP}} \label{sec:thm:omegaIP}

	Suppose $\{t_r\}_{r = 0}^M$ is not an optimal solution of \eqref{eq:IP}.
	Then, there exists a feasible solution $\{t'_r\}_{r = 0}^M$ which can achieve a larger objective value than $\{t_r\}_{r = 0}^M$.
	It means that there exists distinct $m, n \in S$ such that we can reallocate the resource by increasing $t_m$ and decreasing $t_n$ to obtain a larger objective value.

	We use the same $\{t'_r\}_{r = 0}^M$ defined in \eqref{eq:t'_r} to reallocate the resource.
	The increment in the objective value is
	\begin{equation*}
		\sum_{r = 0}^M h_r E_r(t'_r) - \sum_{r = 0}^M h_r E_r(t_r) = s(\Delta_{m, \lfloor t_m \rfloor} - \Delta_{n, \lceil t_n - 1 \rceil}),
	\end{equation*}
	which implies that $\Delta_{m, \lfloor t_m \rfloor} > \Delta_{n, \lceil t_n - 1 \rceil}$.

	Condition i) is violated if $t_m$ is not an integer.
	If $t_m$ is an integer, then the inclusion of $\Delta_{n, \lceil t_n - 1 \rceil}$ and the exclusion of $\Delta_{m, \lfloor t_m \rfloor}$ violate condition ii).
	The proof is done by contraposition.

\section{Proof of Theorem~\ref{thm:intuition}} \label{sec:thm:intuition}

Let $\{t_r\}_{r = 0}^M$ be an optimal solution to \eqref{eq:IP} which has at most one non-integer $t_r$.
Define $\epsilon_m = t_m - \lfloor t_m \rfloor$ and $\epsilon_n = t_n - \lfloor t_n \rfloor$.
Denote $(\alpha_{t|r})_{t = 0}^\infty$ by $\alpha_{r}$ and $(\alpha_r)_{r = 0}^M$ by $\boldsymbol{\alpha}$.

Suppose there exists some $n > m$ such that $t_n < t_m$, where $h_n, h_m \neq 0$.
If we assume $t_r \ge r$ for all $r \in [M]$, then we also have $m < n \le t_n < t_m$.

Let $\boldsymbol{\alpha}$ be the corresponding solution of $\{t_r\}_{r = 0}^M$ for \eqref{eq:P}.
Now, we construct another point $\boldsymbol{\alpha}'$. %
Set $\alpha'_r = \alpha_r$ for $r \neq m, n$.
For $r = m, n$, we partially move the probability masses between $\alpha_{t|m}$ and $\alpha_{t|n}$.
\ifnum\paperversion=1
The definition of $\alpha'_{t|m}$ and $\alpha'_{t|n}$ are shown at the top of the next page.

\begin{figure*}
	\begin{equation*}
		\alpha'_{t|m} = \begin{cases}
			(1-h_n)\epsilon_m & \text{if } t = \lfloor t_m \rfloor + 1,\\
			h_n(1-\epsilon_n) & \text{if } t = \lfloor t_n \rfloor,\\
			\delta_{\lfloor t_n \rfloor + 1, \lfloor t_m \rfloor} h_n \epsilon_n &\\
			\qquad +\; (1-h_n)(1-\epsilon_m) & \text{if } t = \lfloor t_m \rfloor,\\
			\delta_{\lfloor t_n \rfloor + 1, \lfloor t_m \rfloor} (1-h_n)(1-\epsilon_m) &\\
			\qquad +\; h_n \epsilon_n & \text{if } t = \lfloor t_n \rfloor + 1,\\
			0 & \text{otherwise},
		\end{cases}
		\qquad\text{and}\qquad
		\alpha'_{t|n} = \begin{cases}
			h_m \epsilon_m & \text{if } t = \lfloor t_m \rfloor + 1,\\
			(1-h_m)(1-\epsilon_n) & \text{if } t = \lfloor t_n \rfloor,\\
			\delta_{\lfloor t_n \rfloor + 1, \lfloor t_m \rfloor} (1-h_m)\epsilon_n &\\
			\qquad +\; h_m(1-\epsilon_m) & \text{if } t = \lfloor t_m \rfloor,\\
			\delta_{\lfloor t_n \rfloor + 1, \lfloor t_m \rfloor} h_m(1-\epsilon_m) &\\
			\qquad +\; (1-h_m)\epsilon_n & \text{if } t = \lfloor t_n \rfloor + 1,\\
			0 & \text{otherwise}.
		\end{cases}
	\end{equation*}
	\hrule
\end{figure*}
\else
Define
	\begin{equation*}
		\alpha'_{t|m} = \begin{cases}
			(1-h_n)\epsilon_m & \text{if } t = \lfloor t_m \rfloor + 1,\\
			h_n(1-\epsilon_n) & \text{if } t = \lfloor t_n \rfloor,\\
			\delta_{\lfloor t_n \rfloor + 1, \lfloor t_m \rfloor} h_n \epsilon_n + (1-h_n)(1-\epsilon_m) & \text{if } t = \lfloor t_m \rfloor,\\
			\delta_{\lfloor t_n \rfloor + 1, \lfloor t_m \rfloor} (1-h_n)(1-\epsilon_m) + h_n \epsilon_n & \text{if } t = \lfloor t_n \rfloor + 1,\\
			0 & \text{otherwise},
		\end{cases}
	\end{equation*}
and
	\begin{equation*}
		\alpha'_{t|n} = \begin{cases}
			h_m \epsilon_m & \text{if } t = \lfloor t_m \rfloor + 1,\\
			(1-h_m)(1-\epsilon_n) & \text{if } t = \lfloor t_n \rfloor,\\
			\delta_{\lfloor t_n \rfloor + 1, \lfloor t_m \rfloor} (1-h_m)\epsilon_n + h_m(1-\epsilon_m) & \text{if } t = \lfloor t_m \rfloor,\\
			\delta_{\lfloor t_n \rfloor + 1, \lfloor t_m \rfloor} h_m(1-\epsilon_m) + (1-h_m)\epsilon_n & \text{if } t = \lfloor t_n \rfloor + 1,\\
			0 & \text{otherwise}.
		\end{cases}
	\end{equation*}
\fi

The Kronecker delta is introduced to sum up the probability masses which should be assigned to $\lfloor t_m \rfloor$ and $\lfloor t_n + 1 \rfloor$ individually when $\lfloor t_m \rfloor = \lfloor t_n + 1 \rfloor$.

The following shows that $\boldsymbol{\alpha}'$ is a feasible solution to \eqref{eq:P}:
\ifnum\paperversion=1
\begin{IEEEeqnarray*}{Cl}
	& \sum_{r = 0}^M h_r \sum_{t = 0}^\infty t \alpha'_{t|r}\\
	= & \sum_{\substack{r = 0\\r \neq m, n}}^M h_r \sum_{t = 0}^\infty t \alpha_{t|r} + h_m((1-h_n)t_m + h_n t_n)\\
		 & +\; h_n(h_m t_m + (1-h_m)t_n)\\
	= & \sum_{r = 0}^M h_r \sum_{t = 0}^\infty t \alpha_{t|r}\\
	= & t_\text{avg}.
\end{IEEEeqnarray*}
\else
\begin{IEEEeqnarray*}{Cl}
	& \sum_{r = 0}^M h_r \sum_{t = 0}^\infty t \alpha'_{t|r} = \sum_{\substack{r = 0\\r \neq m, n}}^M h_r \sum_{t = 0}^\infty t \alpha_{t|r} + h_m((1-h_n)t_m + h_n t_n) + h_n(h_m t_m + (1-h_m)t_n)\\
	= & \sum_{r = 0}^M h_r \sum_{t = 0}^\infty t \alpha_{t|r} = t_\text{avg}.
\end{IEEEeqnarray*}
\fi

Now, we show that $\boldsymbol{\alpha}'$ can obtain a larger objective value of \eqref{eq:P} than $\boldsymbol{\alpha}$.
Consider
\begin{IEEEeqnarray*}{Cl}
	& \sum_{r = 0}^M h_r \sum_{t = 0}^\infty \alpha'_{t|r} E_r(t) - \sum_{r = 0}^M h_r \sum_{t = 0}^\infty \alpha_{t|r} E_r(t)\\
\ifnum\paperversion=1
	= & h_m((1-h_n) E_m(t_m) + h_n E_m(t_n)) - h_m E_m(t_m)\\
		 & +\; h_n((h_m E_n(t_m) + (1-h_m) E_n(t_n)) - h_n E_n(t_n)\\
	= & h_m h_n (E_m(t_n) - E_m(t_m)) + h_n h_m (E_n(t_m) - E_n(t_n))\\
	= & h_m h_n \sum_{i = \lfloor t_n \rfloor +1}^{\lfloor t_m \rfloor -1} ( \Delta_{n, i} - \Delta_{m, i})\\
		 & \qquad +\; h_m h_n \epsilon_m (\Delta_{n, \lfloor t_m \rfloor} - \Delta_{m, \lfloor t_m \rfloor})\\
		 & \qquad \qquad +\; h_m h_n (1-\epsilon_n) (\Delta_{n, \lfloor t_n \rfloor} - \Delta_{m, \lfloor t_n \rfloor})\\
\else
	= & h_m((1-h_n) E_m(t_m) + h_n E_m(t_n)) - h_m E_m(t_m) + h_n((h_m E_n(t_m) + (1-h_m) E_n(t_n)) - h_n E_n(t_n)\\
	= & h_m h_n (E_m(t_n) - E_m(t_m)) + h_n h_m (E_n(t_m) - E_n(t_n))\\
	= & h_m h_n \sum_{i = \lfloor t_n \rfloor +1}^{\lfloor t_m \rfloor -1} ( \Delta_{n, i} - \Delta_{m, i}) + h_m h_n \epsilon_m (\Delta_{n, \lfloor t_m \rfloor} - \Delta_{m, \lfloor t_m \rfloor}) + h_m h_n (1-\epsilon_n) (\Delta_{n, \lfloor t_n \rfloor} - \Delta_{m, \lfloor t_n \rfloor})\\
\fi
	> & 0,
\end{IEEEeqnarray*}
where the last inequality holds as $\Delta_{n,i} > \ldots > \Delta_{m,i}$ for all $i \ge \lfloor t_n \rfloor$.
That is, $\boldsymbol{\alpha}$ cannot solve \eqref{eq:P}, which contradicts that the solution of \eqref{eq:IP} solves \eqref{eq:P} by Theorem~\ref{thm:IP}.
So, we must have $t_n \ge t_m$.

Now, we assume $\Delta_{r+1,t+1} > \Delta_{r,t}$.
Suppose we have $t_n = t_m > 0$ for some $n > m$ where $h_n, h_m \neq 0$.
Similarly, if we assume $t_r \ge r$ for all $r \in [M]$, then we also have $t_n = t_m \ge n > m$.
We construct $\{t'_r\}_{r = 0}^M$ by
\begin{equation*}
	t'_r = \begin{cases}
		t_m - h_n \sigma & \text{if } r = m,\\
		t_n + h_m \sigma & \text{if } r = n,\\
		t_r & \text{otherwise},
	\end{cases}
\end{equation*}
where 
\begin{equation*}
	\sigma = \begin{cases}
		1 & \text{if } t_n = t_m \text{ are integers,}\\
		\min\{\frac{\epsilon_m}{h_n}, \frac{1-\epsilon_n}{h_m}\}/2 & \text{otherwise.}
	\end{cases}
\end{equation*}

Note that $h_n, h_m \neq 0$ implies that $0 < h_n, h_m < 1$.
So when $t_m = t_n$ are integers, i.e., $t_m \ge 1$, we have $t_m - h_n > 0$ and $t_n + h_m < t_n + 1$.
For the case where $t_m = t_n$ are not integers, we have $h_n \sigma < \epsilon_m$ and $h_m \sigma < 1-\epsilon_n$, so we have $\lfloor t_m - h_n \sigma \rfloor = \lfloor t_m \rfloor$ and $\lfloor t_n + h_m \sigma \rfloor = \lfloor t_n \rfloor$.

It is easy to see that $\{t'_r\}_{r = 0}^M$ is a feasible solution to \eqref{eq:IP}:
\ifnum\paperversion=1
\begin{IEEEeqnarray*}{rCl}
	\sum_{r = 0}^M h_r t'_r & = & \sum_{\substack{r = 0\\r \neq m, n}}^M h_r t_r + h_m(t_m-h_n\sigma) + h_n(t_n+h_m\sigma)\\
   & = & \sum_{r = 0}^M h_r t_r = t_\text{avg}.
\end{IEEEeqnarray*}
\else
\begin{equation*}
	\sum_{r = 0}^M h_r t'_r = \sum_{\substack{r = 0\\r \neq m, n}}^M h_r t_r + h_m(t_m-h_n\sigma) + h_n(t_n+h_m\sigma) = \sum_{r = 0}^M h_r t_r = t_\text{avg}.
\end{equation*}
\fi

To show that $\{t'_r\}$ can obtain a larger objective value of \eqref{eq:IP} than $\{t'_r\}$, we consider the following:
\ifnum\paperversion=1
\begin{IEEEeqnarray*}{Cl}
	& \sum_{r = 0}^M h_r E_r(t'_r) - \sum_{r = 0}^M h_r E_r(t_r)\\
	= & h_m ( E_m(t_m-h_n\sigma) - E_m(t_m) )\\
		 & +\; h_n ( E_n(t_n+h_m\sigma) - E_n(t_n) )\\
	= & h_n h_m \sigma \Delta_{n, \lfloor t_n \rfloor} - h_m h_n \sigma \Delta_{m, \lfloor t_m \rfloor - \delta_{\epsilon_m, 0}}\\
	> & 0.
\end{IEEEeqnarray*}
\else
\begin{IEEEeqnarray*}{Cl}
	& \sum_{r = 0}^M h_r E_r(t'_r) - \sum_{r = 0}^M h_r E_r(t_r) = h_m ( E_m(t_m-h_n\sigma) - E_m(t_m) ) + h_n ( E_n(t_n+h_m\sigma) - E_n(t_n) )\\
	= & h_n h_m \sigma \Delta_{n, \lfloor t_n \rfloor} - h_m h_n \sigma \Delta_{m, \lfloor t_m \rfloor - \delta_{\epsilon_m, 0}} > 0.
\end{IEEEeqnarray*}
\fi
The inequality holds as:
\begin{itemize}
	\item when $\delta_{\epsilon_m, 0} = 0$, we have $\Delta_{n, \lfloor t_n \rfloor} > \ldots > \Delta_{m, \lfloor t_n \rfloor} = \Delta_{m, \lfloor t_m \rfloor}$;
	\item when $\delta_{\epsilon_m, 0} = 1$, we have $\Delta_{n, \lfloor t_n \rfloor} = \Delta_{n, \lfloor t_m \rfloor} > \Delta_{n-1, \lfloor t_m \rfloor-1} \ge \ldots \ge \Delta_{m, \lfloor t_m \rfloor - 1}$.
\end{itemize}
This result contradicts that $\{t_r\}_{r = 0}^M$ solves \eqref{eq:IP}.
Then, we must have $t_m \neq t_n$.

\section{Proof of Corollary~\ref{cor:rhr}} \label{sec:cor:rhr}

	By Corollary~\ref{cor:ratio}, we know that $\Delta_{r,t} = \Delta_{r',t'}$ for all $t < r$ and $t' < r'$.
	By Theorem~\ref{thm:concave}, $E_r(t)$ is concave, i.e., $\Delta_{r,t} \ge \Delta_{r,t+1}$ for all $r, t$.
	So, $\biguplus_{r \in [M]} \Omega_r(r)$ is a collection of the largest $\sum_{r = 0}^M r = M(M+1)/2$ elements in $\biguplus_{r \in [M]} \Omega_r(\infty)$, which consumes $\sum_{r = 0}^M rh_r$ resource.

	When $t_\text{avg} > \sum_{r = 0}^M r h_r$, we have to select more elements into the collection, which means that $t_r \ge r$ for all $r \in [M]$.
	By Theorem~\ref{thm:omegaIP}, there is an optimal solution such that $t_r \ge r$ for all $r \in [M]$.

	Next, if we have $\Pr(Z_1 = 1. Z_{c+1} = 1, \rk(\mathbf{Y}_c) = r-1) \neq 0$ for all $c \ge r$, then by Theorem~\ref{thm:concave}, the concavity of $E_r(t)$ is strict at $c \ge r$ for all $r > 0$.
	This means that $\Delta_{r,t} < \Delta_{r,t-1}$ for all $r > 0$.
	We do not consider $r = 0$ as $E_0(\cdot) = 0$ implies that $\Delta_{0,\cdot} = 0$.
	By Theorem~\ref{thm:IPomega}, we conclude that any optimal solution must satisfy $t_r \ge r$ for all $r \in S$.

\section{Proof of Theorem~\ref{thm:opt}} \label{sec:thm:opt}

	It is trivial that the output is a feasible solution.
	For the second statement, we are having an input that corresponds to a multiset $\mathcal{M} = \biguplus_{r \in [M]} \Omega_r(t_r)$ which is a collection of the largest $\sum_{r \in [M]} t_r = \sum_{r \in [M]} \lceil t_r \rceil$ elements in $\biguplus_{r \in [M]} \Omega_r(\infty)$.
	The algorithm finds and adds the largest element in $\biguplus_{r \in [M]} (\Omega_r(\infty) \setminus \Omega_r(t_r))$ into $\mathcal{M}$, which preserve condition iii) in the definition of a preferred solution.
	Also, the corresponding $t_r$ is increased by $1$ unless there is not enough resource, which only occurs at the last element we add into $\mathcal{M}$ before the algorithm terminates.
	This last element is the smallest element in $\mathcal{M}$, so condition ii) holds.
	Thus, the output is also a preferred solution.

\section{Proof of Theorem~\ref{thm:tune}} \label{sec:thm:tune}

	Let $\max_{r \in [M]} \Delta_{r, \lfloor t_r \rfloor} > \min_{r \in [M]} \Delta_{r, \lceil t_r - 1 \rceil}$.
	The if statements handle the cases that the chosen rank is not in $S$ in a trivial way.
	If both ranks are in $S$, then we can use the same construction of $\{t'_r\}_{r = 0}^M$ in \eqref{eq:t'_r} to increase the objective.

	Note that we have $t'_n \in \mathbb{N}$ or $t'_m \in \mathbb{N}$ (or both).
	The case $t'_n \in \mathbb{N}$ corresponds to the removal of a smallest element from $\biguplus_{r \in S} \Omega_r(t_r)$. %
	Consider the case $t'_m \in \mathbb{N}$.
	If $t_m \in \mathbb{N}$, then it adds a largest element in $\biguplus_{r \in S} (\Omega_r(\infty) \setminus \Omega_r(t_r))$ into $\biguplus_{r \in S} \Omega_r(t_r)$; %
	otherwise it does nothing on the multiset.
	As $\Delta_{m, \lfloor t_m \rfloor} \ge \Delta_{m, \lceil t_m \rceil}$, we know that the loop terminates after all the elements in $\biguplus_{r \in [M]} \Omega_r(t_r)$ are no smaller than those in $\biguplus_{r \in [M]} (\Omega_r(\infty) \setminus \Omega_r(t_r))$, which follows the contraposition of Theorem~\ref{thm:IPomega}.
	Also, only those $r \in \argmin_{r \in [M]} \biguplus_{r \in [M]} \Omega_r(t_r)$ can have non-integer $t_r$, although it may not be unique.

	The algorithm then removes the non-integer parts of $t_r$.
	This step only removes the smallest elements in $\biguplus_{r \in [M]} \Omega_r(t_r)$ when there are non-integer $t_r$.
	So after the removal, the multiset $\biguplus_{r \in [M]} \Omega_r(t_r)$ is a collection of the largest $\sum_{r \in [M]} \lceil t_r \rceil$ elements in $\biguplus_{r \in [M]} \Omega_r(\infty)$ where all $t_r$ are integers.
	By Corollary~\ref{cor:tavg} and Theorem~\ref{thm:omegaIP}, it is a preferred solution of a subproblem having less total resource.
	Then, the algorithm calls Algorithm~\ref{alg:opt}.
	By Theorem~\ref{thm:opt}, the output is a preferred solution of \eqref{eq:IP}.
\section{Discussion on $\Delta \mathbb{E}[\rk(\mathbf{R}_{r+1,t+1})]$ and $\Delta \mathbb{E}[\rk(\mathbf{R}_{r,t})]$} \label{sec:deltart}

In order to compare the difference between $\Delta\mathbb{E}[\rk(\mathbf{R}_{r,t})]$ and $\Delta\mathbb{E}[\rk(\mathbf{R}_{r+1,t+1})]$, we have to formulate their close form formulas.

The close form formula for the probability of the rank of a totally random matrix over finite field can be found in \cite{Landsberg1893,comb}.
However, we only need to use the recursive formula of the above probability.
The technique to formulate the recursive relation can be found in \cite{random_graph}.

Consider $\mathbf{R}_{r,t+1} = (\mathbf{R}'_{r,t} \mid \mathbf{x})$, where $\mathbf{x}$ is a totally random column vector.
Let $V$ be the vector space spanned by the columns in $\mathbf{R}'_{r,t}$.
The vector $\mathbf{x}$ is linearly independent of all the columns in $\mathbf{R}'_{r,t}$ if and only if $\mathbf{x} \not \in V$.
Note that the size of the sample space for $\mathbf{x}$ is $q^r$.
So for $0 \le i \le \min\{r,t\}$, we have
\begin{equation*}
	\Pr(\mathbf{x} \not \in V | \rk(\mathbf{R}'_{r,t}) = i) = 1 - |V|/q^r = 1 - q^{i-r}
\end{equation*}
when $\Pr(\rk(\mathbf{R}'_{r,t}) = i) \neq 0$.
In general, we have
\begin{IEEEeqnarray*}{rCl}
	\Pr(\rk(\mathbf{R}'_{r,t}) = i \wedge \mathbf{x} \not \in V) & = & \Pr(\rk(\mathbf{R}'_{r,t}) = i) (1 - q^{i-r}),\\
	\Pr(\rk(\mathbf{R}'_{r,t}) = i \wedge \mathbf{x} \in V) & = & \Pr(\rk(\mathbf{R}'_{r,t}) = i) q^{i-r},
\end{IEEEeqnarray*}
where we can replace $\mathbf{R}'_{r,t}$ by $\mathbf{R}_{r,t}$ as they are equiprobable.

Next, for $0 \le i \le \min\{r,t+1\}$, we have
\ifnum\paperversion=1
\begin{IEEEeqnarray*}{Cl}
	& \Pr(\rk(\mathbf{R}_{r,t+1}) = i)\\
	= & \Pr(\rk(\mathbf{R}'_{r,t}) = i \wedge \mathbf{x} \in V)\\
	& \qquad +\; \Pr(\rk(\mathbf{R}'_{r,t}) = i-1 \wedge \mathbf{x} \not \in V)\\
	= & \Pr(\rk(\mathbf{R}_{r,t}) = i) q^{i-r}\\
	& \qquad +\; \Pr(\rk(\mathbf{R}_{r,t}) = i-1) (1-q^{i-1-r}). \yesnumber \label{eq:concaveRii}
\end{IEEEeqnarray*}
\else
\begin{IEEEeqnarray*}{rCl}
	\Pr(\rk(\mathbf{R}_{r,t+1}) = i) & = & \Pr(\rk(\mathbf{R}'_{r,t}) = i \wedge \mathbf{x} \in V) + \Pr(\rk(\mathbf{R}'_{r,t}) = i-1 \wedge \mathbf{x} \not \in V)\\
	& = & \Pr(\rk(\mathbf{R}_{r,t}) = i) q^{i-r} + \Pr(\rk(\mathbf{R}_{r,t}) = i-1) (1-q^{i-1-r}). \yesnumber \label{eq:concaveRii}
\end{IEEEeqnarray*}
\fi

Note that we have
\ifnum\paperversion=1
\begin{IEEEeqnarray*}{rCl} 
	\Delta \mathbb{E}[\rk(\mathbf{R}_{r,t})] & = & \sum_{i = 0}^{\min\{r,t\}} \Pr(\rk(\mathbf{R}_{r,t}) = i) (1 - q^{i-r}) \yesnumber \label{eq:concaveRiii} \\
	& = & \Pr(\mathbf{x} \not \in V).
\end{IEEEeqnarray*}
\else
\begin{equation}
	\label{eq:concaveRiii}
	\Delta \mathbb{E}[\rk(\mathbf{R}_{r,t})] = \sum_{i = 0}^{\min\{r,t\}} \Pr(\rk(\mathbf{R}_{r,t}) = i) (1 - q^{i-r}) = \Pr(\mathbf{x} \not \in V).
\end{equation}
\fi

So, we can calculate
\ifnum\paperversion=1
\begin{IEEEeqnarray*}{Cl}
	& \Delta \mathbb{E}[\rk(\mathbf{R}_{r,t+1})] - \Delta \mathbb{E}[\rk(\mathbf{R}_{r,t})]\\
	= & \sum_{i = 0}^{\min\{r,t+1\}} \Pr(\rk(\mathbf{R}_{r,t+1}) = i) (1 - q^{i-r})\\
	& \qquad -\; \sum_{i = 0}^{\min\{r,t\}} \Pr(\rk(\mathbf{R}_{r,t}) = i) (1 - q^{i-r})\\
	= & \sum_{i = -1}^{\min\{r-1,t\}} (1-q^{i+1-r})(1-q^{i-r}) \Pr(\rk(\mathbf{R}_{r,t}) = i)\\
	  & \qquad -\; \sum_{i = 0}^{\min\{r,t\}} (1-q^{i-r})^2 \Pr(\rk(\mathbf{R}_{r,t}) = i) \yesnumber \label{eq:concaveR0} \\
	= & \sum_{i = 0}^{\min\{r,t\}} (1-q^{i-r}) \Pr(\rk(\mathbf{R}_{r,t}) = i) q^{i-r} (1-q), \yesnumber \label{eq:concaveR1} \\
\end{IEEEeqnarray*}
\else
\begin{IEEEeqnarray*}{cl}
	& \Delta \mathbb{E}[\rk(\mathbf{R}_{r,t+1})] - \Delta \mathbb{E}[\rk(\mathbf{R}_{r,t})]\\
	= & \sum_{i = 0}^{\min\{r,t+1\}} \Pr(\rk(\mathbf{R}_{r,t+1}) = i) (1 - q^{i-r}) - \sum_{i = 0}^{\min\{r,t\}} \Pr(\rk(\mathbf{R}_{r,t}) = i) (1 - q^{i-r})\\
	= & \sum_{i = -1}^{\min\{r-1,t\}} (1-q^{i+1-r})(1-q^{i-r}) \Pr(\rk(\mathbf{R}_{r,t}) = i) - \sum_{i = 0}^{\min\{r,t\}} (1-q^{i-r})^2 \Pr(\rk(\mathbf{R}_{r,t}) = i) \yesnumber \label{eq:concaveR0} \\
	= & \sum_{i = 0}^{\min\{r,t\}} (1-q^{i-r}) \Pr(\rk(\mathbf{R}_{r,t}) = i) q^{i-r} (1-q), \yesnumber \label{eq:concaveR1} \\
\end{IEEEeqnarray*}
\fi
where
\begin{itemize}
	\item \eqref{eq:concaveR0} follows \eqref{eq:concaveRii}; and
	\item \eqref{eq:concaveR1} holds since $\Pr(\rk(\mathbf{R}_{r,t}) = -1) = 0$ and $1-q^{i-r} = 0$ when $i = r$, so the sum from $i = -1$ to $\min\{r-1,t\}$ can be replaced by the sum from $i = 0$ to $\min\{r,t\}$.
\end{itemize}

Next, we calculate
\ifnum\paperversion=1
\begin{IEEEeqnarray*}{Cl}
	& \Delta \mathbb{E}[\rk(\mathbf{R}_{r+1,t})]\\
	= & \sum_{i = 0}^{\min\{r+1,t\}} \Pr(\rk(\mathbf{R}_{r+1,t}) = i)(1-q^{i-r-1}) \yesnumber \label{eq:concaveRa} \\
	= & \sum_{i = 0}^{\min\{r+1,t\}} \Pr(\rk(\mathbf{R}_{r,t}) = i)q^{i-t}(1-q^{i-r-1}) \\
	  & +\; \sum_{i = 0}^{\min\{r,t\}} \Pr(\rk(\mathbf{R}_{r,t}) = i)(1-q^{i-t})(1-q^{i-r}) \yesnumber \label{eq:concaveRb} \\
	= &  \Delta \mathbb{E}[\rk(\mathbf{R}_{r,t})]\\
	  & +\; \sum_{i = 0}^{\min\{r,t\}} \Pr(\rk(\mathbf{R}_{r,t}) = i) q^{i-t} (q^{i-r} - q^{i-r-1}), \yesnumber \label{eq:concaveRc} \\
\end{IEEEeqnarray*}
\else
\begin{IEEEeqnarray*}{cl}
	& \Delta \mathbb{E}[\rk(\mathbf{R}_{r+1,t})] = \sum_{i = 0}^{\min\{r+1,t\}} \Pr(\rk(\mathbf{R}_{r+1,t}) = i)(1-q^{i-r-1}) \yesnumber \label{eq:concaveRa} \\
	= & \sum_{i = 0}^{\min\{r+1,t\}} \negthickspace\negthickspace \Pr(\rk(\mathbf{R}_{r,t}) = i)q^{i-t}(1-q^{i-r-1}) + \negthickspace \sum_{i = 0}^{\min\{r,t\}} \Pr(\rk(\mathbf{R}_{r,t}) = i)(1-q^{i-t})(1-q^{i-r}) \yesnumber \label{eq:concaveRb} \\
	= &  \Delta \mathbb{E}[\rk(\mathbf{R}_{r,t})] + \sum_{i = 0}^{\min\{r,t\}} \Pr(\rk(\mathbf{R}_{r,t}) = i) q^{i-t} (q^{i-r} - q^{i-r-1}), \yesnumber \label{eq:concaveRc} \\
\end{IEEEeqnarray*}
\fi
where
\begin{itemize}
	\item \eqref{eq:concaveRa} follows \eqref{eq:concaveRiii};
	\item \eqref{eq:concaveRb} holds since the ranks of a matrix and its transpose are equal, so we can apply \eqref{eq:concaveRii} by swapping $r$ and $t$; and
	\item \eqref{eq:concaveRc} holds as $\Pr(\rk(\mathbf{R}_{r,t}) = r+1) = 0$ and by \eqref{eq:concaveRiii}.
\end{itemize}

Now, we can compare the difference between $\Delta\mathbb{E}[\rk(\mathbf{R}_{r,t})]$ and $\Delta\mathbb{E}[\rk(\mathbf{R}_{r+1,t+1})]$.
\ifnum\paperversion=1
We consider the calculation at the top of the next page, where
\else
	\begin{IEEEeqnarray*}{Cl}
		& \Delta\mathbb{E}[\rk(\mathbf{R}_{r+1,t+1})] - \Delta\mathbb{E}[\rk(\mathbf{R}_{r,t})]\\
		= & \Delta\mathbb{E}[\rk(\mathbf{R}_{r+1,t+1})] - \Delta\mathbb{E}[\rk(\mathbf{R}_{r,t+1})] + \Delta\mathbb{E}[\rk(\mathbf{R}_{r,t+1})] - \Delta\mathbb{E}[\rk(\mathbf{R}_{r,t})]\\
		= & \sum_{i = 0}^{\min\{r,t+1\}} \Pr(\rk(\mathbf{R}_{r,t+1}) = i) q^{2i-t-r-2} (q-1)\\
		& \qquad + \sum_{i = 0}^{\min\{r,t\}} \Pr(\rk(\mathbf{R}_{r,t}) = i) (q^{i-r} - 1) q^{i-r} (q-1) \yesnumber \label{eq:diagR1} \\
		= & \sum_{i = 0}^{\min\{r,t\}} \Pr(\rk(\mathbf{R}_{r,t}) = i) q^{i-r} (q-1) (q^{2i-t-r-2} + q^{i-r} - 1)\\
		& \qquad + \sum_{i = 0}^{\min\{r,t\}} \Pr(\rk(\mathbf{R}_{r,t}) = i) (1 - q^{i-r}) q^{2i-t-r} (q-1) \yesnumber \label{eq:diagR2} \\
		= & \sum_{i = 0}^{\min\{r,t\}} \Pr(\rk(\mathbf{R}_{r,t}) = i) q^{i-r} (q-1) (q^{i-t} q^{i-r}  q^{-2} - (q^{i-t}-1)(q^{i-r}-1)). \yesnumber \label{eq:diagR3}
	\end{IEEEeqnarray*}
\fi
\begin{itemize}
	\item \eqref{eq:diagR1} follows \eqref{eq:concaveR1} and \eqref{eq:concaveRc}; and
	\item \eqref{eq:diagR2} follows \eqref{eq:concaveRii} and the fact that $1 - q^{i-r} = 0$ when $i = r$.
\end{itemize}

\ifnum\paperversion=1
\begin{figure*}
	\begin{IEEEeqnarray*}{Cl}
		& \Delta\mathbb{E}[\rk(\mathbf{R}_{r+1,t+1})] - \Delta\mathbb{E}[\rk(\mathbf{R}_{r,t})] = \Delta\mathbb{E}[\rk(\mathbf{R}_{r+1,t+1})] - \Delta\mathbb{E}[\rk(\mathbf{R}_{r,t+1})] + \Delta\mathbb{E}[\rk(\mathbf{R}_{r,t+1})] - \Delta\mathbb{E}[\rk(\mathbf{R}_{r,t})]\\
		= & \sum_{i = 0}^{\min\{r,t+1\}} \Pr(\rk(\mathbf{R}_{r,t+1}) = i) q^{2i-t-r-2} (q-1) + \sum_{i = 0}^{\min\{r,t\}} \Pr(\rk(\mathbf{R}_{r,t}) = i) (q^{i-r} - 1) q^{i-r} (q-1) \yesnumber \label{eq:diagR1} \\
		= & \sum_{i = 0}^{\min\{r,t\}} \Pr(\rk(\mathbf{R}_{r,t}) = i) q^{i-r} (q-1) (q^{2i-t-r-2} + q^{i-r} - 1) + \sum_{i = 0}^{\min\{r,t\}} \Pr(\rk(\mathbf{R}_{r,t}) = i) (1 - q^{i-r}) q^{2i-t-r} (q-1) \yesnumber \label{eq:diagR2} \\
		= & \sum_{i = 0}^{\min\{r,t\}} \Pr(\rk(\mathbf{R}_{r,t}) = i) q^{i-r} (q-1) (q^{i-t} q^{i-r}  q^{-2} - (q^{i-t}-1)(q^{i-r}-1)). \yesnumber \label{eq:diagR3}
	\end{IEEEeqnarray*}
	\hrule
\end{figure*}
\fi

The summand in \eqref{eq:diagR3} is not always non-negative for all $i \in [\min\{r,t\}]$, which makes it difficult to conclude whether the overall sum is non-negative or not.
The sign of each summand is the sign of the term
\begin{equation} \label{eq:deltaneg}
	q^{i-t} q^{i-r}  q^{-2} - (q^{i-t}-1)(q^{i-r}-1).
\end{equation}
To see that \eqref{eq:deltaneg} can be negative, we substitute $i = 0$.
Then, \eqref{eq:deltaneg} is negative if and only if $(q^r-1)(q^t-1) > q^{-2}$, which is true for all $r, t > 0$.

\end{document}